\documentclass[12pt,english,floatfix,nofootinbib,superscriptaddress,aps,prd,preprint]{revtex4}
\usepackage[utf8]{inputenc}       
\usepackage[english]{babel}       
\usepackage[utf8]{inputenc}
\usepackage{textcomp}
\usepackage{amsmath,amsthm,amssymb,amsfonts,mathrsfs,amsbsy} 
\usepackage{tensor}               
\usepackage{slashed}  
\usepackage{bbm}
\usepackage{braket}
\usepackage{esint}                
\usepackage{cancel}               
\usepackage{graphicx}             
\usepackage{natbib}
\usepackage{float}                
\usepackage{subfig}               
\usepackage[font=small,labelfont=bf]{caption} 
\usepackage{multirow}             
\usepackage{array}                
\usepackage{tikz}                 
\usetikzlibrary{quotes,angles,arrows,decorations.markings} 
\usepackage{makecell} 

\usepackage{lipsum}

\usepackage{xcolor}               
\usepackage{color}                
\usepackage{textcomp}             
\usepackage{units}                

\newcommand{\be}{\begin{equation}}
\newcommand{\ee}{\end{equation}}
\newcommand{\Be}{\begin{eqnarray}}
\newcommand{\Ee}{\end{eqnarray}}

\newcommand{\mincir}{\raise
-3.truept\hbox{\rlap{\hbox{$\sim$}}\raise4.truept\hbox{$<$}\ }}
\newcommand{\magcir}{\raise
-3.truept\hbox{\rlap{\hbox{$\sim$}}\raise4.truept\hbox{$>$}\ }}

\newcolumntype{Y}{>{\centering\arraybackslash}X}
\providecommand{\U}[1]

\usepackage[dvips]{epsfig}        
\usepackage[dvips]{graphicx}      
\usepackage{hyperref}             
\hypersetup{
    colorlinks=true,              
    breaklinks=true,              
    citecolor=blue,               
    linkcolor=[rgb]{0,0.5,0.9},   
    urlcolor=red,                 
    filecolor=green               
}

\newcommand{\ie}{\begin{equation}}
\newcommand{\fe}{\end{equation}}
\newcommand{\se}{\begin{eqnarray}}
\newcommand{\ff}{\end{eqnarray}}

\begin{document}

\title{Neutrino oscillations induced by a new bumblebee black hole}


\author{Yuxuan Shi}
\email{shiyx2280771974@gmail.com}
\affiliation{Department of Physics, East China University of Science and Technology, Shanghai 200237, China}


\author{A. A. Ara\'{u}jo Filho}
\email{dilto@fisica.ufc.br}
\affiliation{Departamento de Física, Universidade Federal da Paraíba, Caixa Postal 5008, 58051--970, João Pessoa, Paraíba,  Brazil.}
\affiliation{Departamento de Física, Universidade Federal de Campina Grande Caixa Postal 10071, 58429-900 Campina Grande, Paraíba, Brazil.}
\affiliation{Center for Theoretical Physics, Khazar University, 41 Mehseti Street, Baku, AZ-1096, Azerbaijan.}


\date{\today}

\begin{abstract}

This work investigates neutrino propagation in the spacetime of a newly introduced black hole arising from spontaneous Lorentz--symmetry breaking in bumblebee gravity. The analysis focuses on three independent components: the rate at which neutrino–antineutrino annihilation deposits energy in the surrounding region, the geometric contribution to the phase accumulated by neutrino mass eigenstates, and the modifications to flavor conversion produced by weak gravitational lensing. Working with a two--flavor system, both mass orderings are examined, and the calculation incorporates the interference between distinct trajectories reaching the detector. The numerical results show that the Lorentz--violating deformation substantially increases the efficiency of the annihilation channel, produces characteristic shifts in the oscillation phase not present in earlier bumblebee configurations, and reshapes the angular dependence of the lensing--induced flavor transition pattern.

\end{abstract}


\maketitle

\tableofcontents


\section{Introduction }

The idea that local Lorentz symmetry might operate only approximately, rather than as a strict principle of nature, has gained momentum in light of approaches to quantum gravity that foresee corrections to spacetime structure at accessible energy scales \cite{kostelecky1989spontaneous,colladay1997cpt,kostelecky2004gravity,kostelecky1999constraints,kostelecky2011data}. In several scenarios, such deviations arise when certain fields stabilize at nonzero vacuum values, thereby selecting a directional background and inducing spontaneous breaking of Lorentz invariance.
A compact and effective framework that encapsulates this mechanism is provided by bumblebee models. In these theories, a vector field is constrained by a potential to maintain a fixed magnitude, and once the vacuum configuration is reached, the geometry acquires a preferred direction. The resulting spacetime incorporates controlled departures from standard general relativity in a manner that remains consistent and theoretically well defined \cite{Bluhm:2023kph,bluhm2005spontaneous,Bluhm:2019ato,Maluf:2014dpa,bluhm2008spontaneous,Maluf:2013nva}.

Bumblebee constructions have been repeatedly employed as effective frameworks for studying departures from exact Lorentz invariance. These theories arise naturally in several contexts, ranging from mechanisms inspired by string theory that allow for spontaneous symmetry breaking \cite{kostelecky1991photon,kostelecky1989spontaneous} to vector–tensor modifications of Einstein’s equations \cite{jacobson2004einstein} and broader effective field–theory treatments in which background fields acquire nontrivial configurations \cite{bluhm2005spontaneous,kostelecky2004gravity}. At the core of these models lies a vector field $B_{\mu}$ whose dynamics are governed by a potential enforcing a fixed magnitude, typically expressed as $V(B_{\mu}B^{\mu}\mp b^{2})$. When this condition is satisfied, the field settles into a vacuum configuration that selects a spacetime direction and spontaneously breaks Lorentz symmetry \cite{bluhm2008spontaneous,bluhm2005spontaneous}. This mechanism reorganizes the perturbative sector: fluctuations along the symmetry–preserving directions behave as Nambu–Goldstone modes with photon–like character \cite{bluhm2005spontaneous}, whereas excitations orthogonal to the fixed norm acquire mass through the same potential \cite{bluhm2008spontaneous}.

Formulations of the bumblebee mechanism have been extended to curved spacetimes, where the fixed-norm vacuum configuration of the vector field becomes intertwined with the gravitational dynamics \cite{Bertolami:2005bh}. After this coupling to geometry was established, the framework branched into several distinct research programs. In the context of compact objects, the solution presented in \cite{Casana:2017jkc} provided a foundation for examining quantum processes near black hole horizons, ranging from modifications of entanglement behavior \cite{Liu:2024wpa} to analyses of particle creation in Lorentz--violating backgrounds \cite{AraujoFilho:2025hkm} (and the corresponding generalization based on tensor-field configurations in the Kalb--Ramond framework \cite{AraujoFilho:2024ctw}). Other investigations shifted toward cosmology and astrophysics. Models featuring anisotropic expansion reminiscent of Kasner--type geometries were proposed in \cite{Neves:2022qyb}, and subsequent work addressed stellar configurations within the same Lorentz--violating framework \cite{Neves:2024ggn}. The influence of the vector background on gravitational--wave propagation was also explored, revealing characteristic deviations from general relativity \cite{Liang:2022hxd,amarilo2024gravitational}.
Further extensions modified the geometric sector itself. Some analyses incorporated a cosmological constant into the bumblebee setup \cite{Maluf:2020kgf}, while others constructed axisymmetric backgrounds through a refined Newman--Janis procedure \cite{Kumar:2025bim}.

Several extensions of bumblebee gravity have been developed in recent years, spanning frameworks far more general than the original static solution of Ref.~\cite{Casana:2017jkc}. Work in the \textit{metric--affine} formulation produced new families of geometries: a static configuration was constructed in \cite{Filho:2022yrk}, and an axisymmetric background was later obtained in \cite{AraujoFilho:2024ykw}. The same setting subsequently allowed for the formulation of a non--commutative counterpart \cite{AraujoFilho:2025rvn} (and the corresponding extension formulated within Kalb--Ramond gravity \cite{AraujoFilho:2025jcu}), enlarging the catalogue of spacetimes compatible with spontaneous Lorentz violation. Research has also moved well beyond black holes. A substantial body of literature now investigates situations where the fixed--norm vector field sustains wormhole structures or preserves their traversability \cite{Magalhaes:2025nql,AraujoFilho:2024iox,Magalhaes:2025lti,Ovgun:2018xys}. In related directions, black--bounce geometries supported by $\kappa$--essence fields have been formulated within symmetry-breaking environments \cite{Pereira:2025xnw}. Propagation effects have been explored as well. Neutrino lensing, for instance, has been studied in purely metric realizations of the theory \cite{Shi:2025plr}, in \textit{metric--affine} versions \cite{Shi:2025ywa}, and in tensor-based generalizations of the bumblebee sector \cite{Shi:2025rfq}.

Neutrinos occupy a distinctive position among elementary particles because their extremely small masses and the mismatch between interaction and propagation bases enable coherent evolution over very long distances—from laboratory scales to astrophysical baselines \cite{Pontecorvo2,Pontecorvo1,maki1962remarks}. This capacity for maintaining quantum coherence makes them highly responsive to physics beyond the Standard Model and to potential gravitational modifications \cite{neu44,neu42,neu43}. In the weak sector, neutrinos are created and detected in flavor configurations, yet their propagation is governed by states of definite mass. Because these two bases do not align, a neutrino produced with a given flavor effectively travels as a superposition of different mass eigenstates. Interference among these components gradually reshapes the flavor content during propagation \cite{neu40,neu41,neu39}. The continual reshuffling of the flavor mixture yields the phenomenon known as neutrino oscillation.

In the standard treatment of neutrino propagation in flat spacetime, the relevant quantities are not the individual masses but the separations between their squared values. The oscillatory behavior hinges on parameters of the form
$\Delta m^{2}_{ij}=m_i^{2}-m_j^{2}$,
which encode the relative spacing among the mass eigenstates. Analyses typically focus on the magnitudes of $\Delta m_{21}^{2}$, $\Delta m_{31}^{2}$, and $\Delta m_{23}^{2}$. Because transition probabilities depend solely on these differences, oscillation experiments reveal only the pattern of mass splittings and remain blind to the absolute neutrino-mass scale \cite{neu45}.

Neutrino flavor evolution acquires additional structure when the particles travel through curved backgrounds. Unlike the flat--spacetime case, where only the mass–squared splittings govern the oscillation pattern, propagation in a gravitational field modifies the phase accumulated by each mass eigenstate \cite{Shi:2025rfq,Chakrabarty:2023kld,Shi:2024flw,AraujoFilho:2025rzh,Alloqulov:2024sns}. These geometric contributions introduce dependence on parameters that are normally inaccessible, including possible sensitivity to the absolute mass scale. The impact of curvature becomes especially relevant for neutrinos originating from energetic astrophysical or cosmological environments, where long baselines and strong gravitational potentials enhance the accumulated phase shift \cite{Shi:2023kid}. As the waves traverse regions with nontrivial geometry, the altered phase relations reflect both the properties of the medium and the internal characteristics of the mass eigenstates. Consequently, deviations between observed flavor ratios and those forecast by standard flat–space oscillation formulas may signal gravitationally induced modifications. This strategy has been explored as a means to probe the neutrino mass spectrum while simultaneously examining the gravitational contexts through which the particles propagate \cite{neu49,neu53,neu47,neu46,neu48,neu51,neu50,neu52,Shi:2023hbw}.

The evolution of neutrino flavors changes noticeably when their trajectories unfold in a curved spacetime. In such a description, the accumulated phase is influenced by the underlying geometry rather than depending solely on intrinsic particle parameters \cite{neu54}. Regions dominated by strong gravitational fields—particularly those surrounding compact objects—can deflect neutrino paths through lensing effects. When these trajectories bend or intersect, the relative phases of the mass eigenstates are modified, reshaping the interference pattern that governs flavor conversion. As a result, the transition probabilities deviate from their flat–space expectations \cite{Shi:2025rfq,Shi:2024flw,neu53}.

The role of gravity in shaping neutrino coherence has been explored in increasing detail, with particular attention to how curved geometries modify the conditions required for flavor conversion \cite{neu57,neu56,neu58}. When the gravitational environment includes rotation, the situation becomes more intricate. Angular momentum of the central object contributes additional terms to the neutrino phase, altering the propagation in ways that depend on the spin of the source. Swami’s analysis demonstrated that these rotational contributions may either diminish or enhance the transition probabilities, depending on the specific features of the rotating spacetime. Such effects can be significant even around stellar--mass objects, placing rotating backgrounds among the relevant arenas for studying gravitational impacts on oscillation behavior \cite{neu59}.

Studies of neutrino oscillations have also been carried out in backgrounds that break spherical symmetry. Axially distorted gravitational fields, often characterized by a deformation parameter $\gamma$, reshape the form of static and asymptotically flat metrics. When neutrinos move through such geometries, the altered spacetime structure modifies the phase accumulated by the mass eigenstates. This modification can produce a sensitivity to the absolute values of the neutrino masses—an effect absent in the standard flat–space formulation—thereby emphasizing how geometric departures from spherical symmetry influence the oscillation pattern \cite{neu60}.

New black hole configurations generated within the bumblebee framework have recently been constructed in Refs.~\cite{Liu:2025oho,Zhu:2025fiy}. Their static form, displayed in Eq.~(\ref{newblacbumblebee}), was later examined in detail in Ref.~\cite{AraujoFilho:2025zaj}, where various gravitational features were investigated, including quasinormal spectra, time--domain evolution, geodesic structure, shadows, lensing properties, topological aspects, and Solar System bounds, together with the behavior of massive particles. Although these studies explored a wide range of classical and semiclassical characteristics, the implications of this bumblebee black hole for neutrino propagation have not yet been addressed. The present work focuses precisely on this missing sector. Our analysis is organized around three main themes: the modification of the neutrino–antineutrino annihilation energy deposition rate, the changes introduced in the oscillation phase by the background geometry, and the effects of gravitational deflection on flavor--conversion probabilities.


\section{Geometry of the new bumblebee black hole }

A recently introduced black hole geometry emerging from the bumblebee framework has been formulated in Refs.~\cite{Liu:2025oho,Zhu:2025fiy}, where the spontaneous breaking of Lorentz symmetry shapes the resulting spacetime, as follows
\ie
\label{newblacbumblebee}
\mathrm{d}s^{2} = - \frac{1}{1+\chi}\left(1 - \frac{2M}{r}         \right)\mathrm{d}t^{2} + \frac{1+\chi}{\left(1 - \frac{2M}{r} \right)} \mathrm{d}r^{2} + r^{2}\mathrm{d}\Omega^{2}.
\fe
In this geometry, the parameter governing Lorentz violation is constructed from the combination $\ell \equiv \xi\,\tilde b^{2}$, where $\xi$ is the coupling constant and $\tilde b^{2}=\tilde b_{\mu}\tilde b^{\mu}$ denotes the vacuum value of the bumblebee field. An additional integration constant, $\tilde\alpha$, appears in the definition of the deformation parameter through the relation $\chi \equiv \tilde\alpha\,\ell$.
A comprehensive analysis of the spacetime in Eq.~(\ref{newblacbumblebee}) was recently carried out in Ref.~\cite{AraujoFilho:2025zaj}. That work examined several gravitational aspects of the solution, including the structure of quasinormal spectra, numerical time--domain evolution, properties of geodesic motion, the associated shadows, lensing behavior, topological characteristics, the dynamics of massive probes, and constraints derived from Solar System observations. Despite this broad range of investigations, no dedicated study has yet addressed how this particular bumblebee black hole influences neutrino oscillations. The following sections are therefore aimed at filling this gap.


\section{Neutrino-Antineutrino Annihilation and Energy Deposition }

This part of the analysis examines how energy is deposited in the gravitational environment governed by the Lorentz--violating parameter $\chi$, defined through Eq.~(\ref{newblacbumblebee}). In this setting, the dominant contribution to the local energy transfer arises from neutrino--antineutrino annihilation. The corresponding deposition rate, expressed per unit volume and per unit time, is described by the formulation presented in Ref.~\cite{Salmonson:1999es}:
\begin{align}
\dfrac{\mathrm{d}\mathrm{E}(r)}{\mathrm{d}t\mathrm{d}V}=2 \, \mathrm{K} \,\mathrm{G}_{f}^{2}\, f(r)\iint
n(\varepsilon_{\nu})n(\varepsilon_{\overline{\nu}})
(\varepsilon_{\nu} + \varepsilon_{\overline{\nu}})
\varepsilon_{\nu}^{3}\varepsilon_{\overline{\nu}}^{3}
\mathrm{d}\varepsilon_{\nu}\mathrm{d} \varepsilon_{\overline{\nu}}
\end{align}
in which
\begin{align}
\mathrm{K} = \dfrac{1}{6\pi}(1\pm4\sin^{2}\vartheta_{W}+8\sin^{4} \vartheta_{W}).
\end{align}

Using the usual numerical choice for the Weinberg angle, $\sin^{2}\vartheta_{W}=0.23$, one can express the contribution of each neutrino--antineutrino channel to the local energy release in closed form. The expressions reported in Ref.~\cite{Salmonson:1999es} show how the weak couplings and the specific flavor content of the interacting pair shape the resulting energy--deposition rate
\begin{equation}
\mathrm{K}(\nu_{\mu},\overline{\nu}_{\mu}) = \mathrm{K}(\nu_{\tau},\overline{\nu}_{\tau})
=\dfrac{1}{6\pi}\left(1-4\sin^{2}\vartheta_{W} + 8\sin^{4}\vartheta_{W}\right),
\end{equation}
and also we have
\begin{equation}
\mathrm{K}(\nu_{e},\overline{\nu}_{e})
=\dfrac{1}{6\pi}\left(1+4\sin^{2}\vartheta_{W} + 8\sin^{4}\vartheta_{W}\right).
\end{equation}

The determination of the energy released through each neutrino--antineutrino annihilation channel depends on the weak--interaction parameters used as input. Adopting the standard choice $\sin^{2}\vartheta_{W}=0.23$ fixes the relative couplings among the participating flavors, while the interaction amplitude is set by the Fermi constant, taken as $\mathrm{G}_{f}=5.29\times10^{-44}\,\text{cm}^{2}\,\text{MeV}^{-2}$. With these quantities specified, one can evaluate the individual contributions of the different flavor pairs. Carrying out the angular integration leads to the deposition formula reported in Ref.~\cite{Salmonson:1999es}
\begin{align}
f(r)&=\iint\left(1-\bm{\Omega_{\nu}}\cdot\bm{\Omega_{\overline{\nu}}}\right)^{2}
\mathrm{d}\Omega_{\nu}\mathrm{d}\Omega_{\overline{\nu}}\notag =\dfrac{2\pi^{2}}{3}(1 - x)^{4}\left(x^{2} + 4x + 5\right)
\end{align}
with $x = \sin\vartheta_{r}$.

For a fixed radial position $r$, the quantity $\vartheta_{r}$ characterizes the inclination of a particle’s path relative to the tangential direction of a circular trajectory at that same radius. The propagation of neutrinos and antineutrinos is described through the unit vectors $\Omega_{\nu}$ and $\Omega_{\overline{\nu}}$, and the integration over their possible directions is carried out using the corresponding solid--angle elements $\mathrm{d}\Omega_{\nu}$ and $\mathrm{d}\Omega_{\overline{\nu}}$. When the system is assumed to be in thermal equilibrium at temperature $T$, the occupation numbers for neutrinos and antineutrinos, $n(\varepsilon_{\nu})$ and $n(\varepsilon_{\overline{\nu}})$, are given by the usual Fermi--Dirac distributions \cite{Salmonson:1999es}
\begin{align}
n(\varepsilon_{\nu}) = \frac{2}{h^{3}}\dfrac{1}{\exp{\left({\frac{\varepsilon_{\nu}}{kT}}\right)} + 1}.
\end{align}

The evaluation of the energy released through neutrino--antineutrino annihilation relies on a framework that incorporates the fundamental constants of quantum and thermal physics, namely Planck’s constant $h$ and Boltzmann’s constant $k$. Once these constants are introduced, the expression for the energy--deposition rate—defined per unit time and per unit volume—follows directly. This formulation captures the quantum--statistical nature of the process and is presented in the analysis of Ref.~\cite{Salmonson:1999es}
\ie
\frac{\mathrm{d}\mathrm{E}}{\mathrm{d}t\mathrm{d}V} = \frac{21\zeta(5)\pi^{4}}{h^{6}}\mathrm{K} \, \mathrm{G}_{f}^{2} \, f(r)(k \, \Tilde{T}(r))^{9}.
\fe
The quantity $\mathrm{d}E/(\mathrm{d}t \,\mathrm{d}V)$ provides a measure of how efficiently energy is generated or redistributed within a localized region surrounding a compact object \cite{Salmonson:1999es}. Its value reflects the radial dependence of the relevant physical conditions, making the temperature distribution $\Tilde{T}(r)$ a fundamental ingredient in describing the thermal state of the environment through which neutrinos and antineutrinos propagate \cite{Salmonson:1999es}.

For an observer situated at radius $r$, the temperature inferred locally is altered by gravitational redshift. This effect imposes the condition $
\Tilde{T}(r)\,\sqrt{-g_{tt}(r,\chi)}=\text{constant}$,
which reflects the role of spacetime curvature in determining thermal measurements \cite{Salmonson:1999es}. When evaluated at the neutrinosphere—located at $r=R$—the emission temperature must satisfy the corresponding redshifted relation presented in Ref.~\cite{Salmonson:1999es}:
\ie
\Tilde{T}(r)\sqrt{-g_{tt}(r,\chi)} = \Tilde{T}(R)\sqrt{-g_{tt}(r,\chi)}.
\fe  
The radius $R$ specifies the location of the neutrinosphere, which acts as the emitting surface for the compact object generating the gravitational field. Using the redshift relation established earlier, the temperature profile $\Tilde{T}(r)$ can be recast in a form more suitable for calculations. Incorporating this redshifted temperature into the energy–transport framework yields the expression for the neutrino luminosity presented in Ref.~\cite{Salmonson:1999es}:
\ie
\mathfrak{L}_{\infty} = -g_{tt}(r,\chi)L(R).
\fe

The luminosity generated by an individual neutrino species at the neutrinosphere follows directly from the emission conditions imposed at $r=R$. Using the corresponding redshifted temperature, the contribution from each flavor is obtained from the expression given in Ref.~\cite{Salmonson:1999es}:
\ie
\mathfrak{L}(R) = 4 \pi R_{0}^{2}\dfrac{7}{4}\dfrac{a\,c}{4}\Tilde{T}^{4}(R).
\fe

In this expression, $c$ represents the speed of light in vacuum, while $a$ is the constant governing black--body radiation. The temperature measured by an observer at radius $r$ must incorporate the influence of gravitational redshift. When this effect is included, the local temperature becomes tied to the metric component $g_{tt}(r,\chi)$, leading to the relation \cite{Salmonson:1999es}:
\ie
\begin{split}
\frac{\mathrm{d}\mathrm{E}(r)}{\mathrm{d}t \, \mathrm{d}V} & = \dfrac{21\zeta(5)\pi^{4}}{h^{6}}
\mathrm{K} \, \mathrm{G}_{f}^{2} \, k^{9}\left(\dfrac{7}{4}\pi a\,c\right)^{-9/4}\mathfrak{L}_{\infty}^{9/4}f(r)
\left[\dfrac{\sqrt{-g_{tt}(r,\chi)}}{-g_{tt}(r,\chi)}\right]^{9/2} R^{-9/2}.
\end{split}
\fe

The symbol $\zeta(s)$ appearing in the expression corresponds to the Riemann zeta function. For real values of $s$ exceeding unity, this function is introduced through the convergent series
\begin{equation}
\zeta(s)=\sum_{n=1}^{\infty} n^{-s}.
\end{equation}
The local energy–deposition rate is influenced both by the radial position and by the geometric properties of the spacetime, with the metric evaluated at the compact object’s surface playing a central role. Obtaining the total energy output in a curved background therefore requires integrating this local rate over the relevant time interval.
The angular part of the calculation, represented by 
$f(r)$, involves a careful treatment of the variable $x$, which was earlier introduced in connection with the angular integration. The refinement of this variable proceeds through the following steps:\cite{Salmonson:1999es,Shi:2023kid,AraujoFilho:2024mvz,Lambiase:2020iul}:
\begin{align}
x^{2}& = \sin^{2}\vartheta_{r}|_{\vartheta_{R}=0}\notag\\
&=1-\dfrac{R^{2}}{r^{2}}\dfrac{g_{tt}(r,\chi)}{g_{tt}(r,\chi)}.
\end{align}

Determining the total energy deposited outside the compact object requires integrating the local deposition rate—given per unit time and per unit volume—over the spatial domain influenced by the gravitational field. In this procedure, the angular factor plays a essential role, and its form is fixed entirely by the background metric. The geometry of the spacetime therefore dictates how this angular contribution behaves throughout the region of interest \cite{Shi:2023kid,Lambiase:2020iul,AraujoFilho:2025rzh,Shi:2025rfq}
\begin{align}
\dot{Q} & = \frac{\mathrm{d}\mathrm{E}}{\sqrt{-g_{tt}(r,\chi)}\mathrm{d}t}\\
&=\dfrac{84\zeta(5)\pi^{5}}{h^{6}}\mathrm{K}\, \mathrm{G}_{f}^{2} \, k^{9}
\left(\dfrac{7}{4}\pi a\,c\right)^{-9/4}
\mathfrak{L}_{\infty}^{9/4}\left[-g_{tt}(r,\chi)\right]^{9/4}\\
& \times 
R^{-3/2}\int_{1}^{\infty}(x-1)^4\left(x^2+4x+5\right)\sqrt{\dfrac{g_{rr}(yR,\chi)}{-g_{tt}^9(yR,\chi)}}y^2\mathrm{d}y.
\end{align}

Here, the quantity $ \dot{Q} $ is used to express the net rate at which neutrino interactions supply energy in the form of electron--positron pairs at a specified radius \cite{Salmonson:1999es}. If this rate grows large enough, the produced pairs can act as a driver for various high--energy astrophysical phenomena. To understand the role played by gravity in this process, one may place the relativistic expression for $ \dot{Q} $ alongside the corresponding Newtonian estimate. This comparison highlights how curved spacetime alters the overall effectiveness of neutrino--powered energy deposition \cite{Salmonson:1999es,Lambiase:2020iul,Shi:2023kid,shi2022neutrino}
\ie
\begin{split}
\label{ratio_Q}
 \frac{\dot{Q}}{\dot{Q}_{Newt}} = 3\left[-g_{tt}(r,\chi)\right]^{9/4}
& \int_{1}^{\infty}(x - 1)^{4}\left(x^{2} + 4x + 5\right)\sqrt{\dfrac{g_{rr}(yR,\chi)}{-g_{tt}^9(yR,\chi)}}y^2\mathrm{d}y.
\end{split}
\fe
where we also have
\ie
\begin{split}
g_{tt}(r,\chi)&= -\dfrac{1}{1+\chi}\left(1 - \frac{2M}{R}\right),\\
g_{tt}(yR,\chi)&= -\dfrac{1}{1+\chi}\left(1 - \frac{2M}{yR}\right).
\end{split}
\fe
In addition, we get
\begin{align}
x^{2}=1-\dfrac{1}{y^{2}}\frac{1-\frac{2M}{yR}}{1-\frac{2M}{R}}.
\end{align}

\begin{figure*}
\centering
\includegraphics[height=6.5cm]{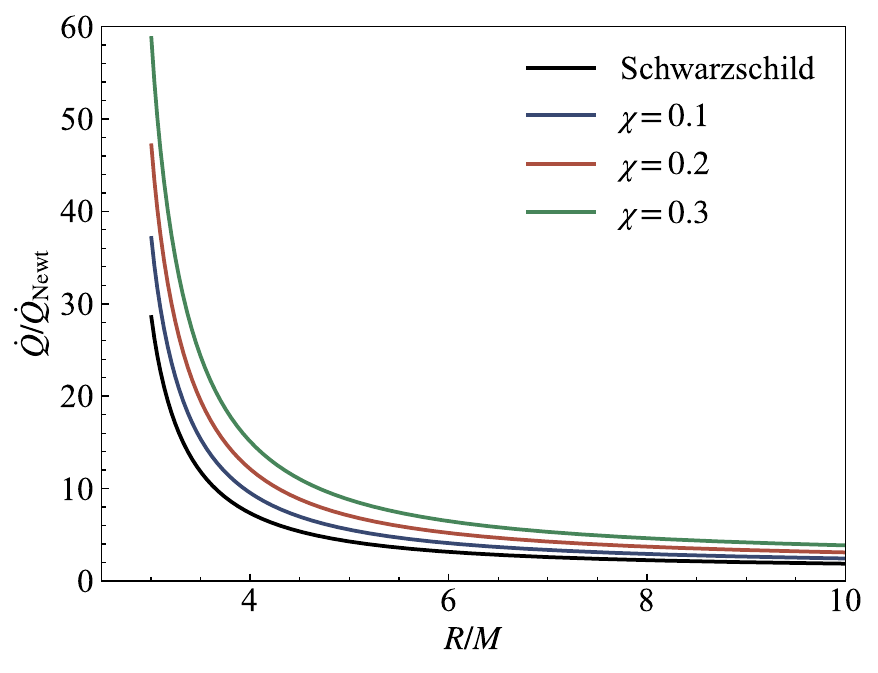}
\caption{Behavior of the ratio, $\dot{Q}/\dot{Q}_{Newt}$, as a function of $R/M$ for different choices of the Lorentz--violating parameter $\chi$.}
\label{fig:neu_FMR_1}
\end{figure*}

\begin{figure*}
\centering
\includegraphics[height=6.5cm]{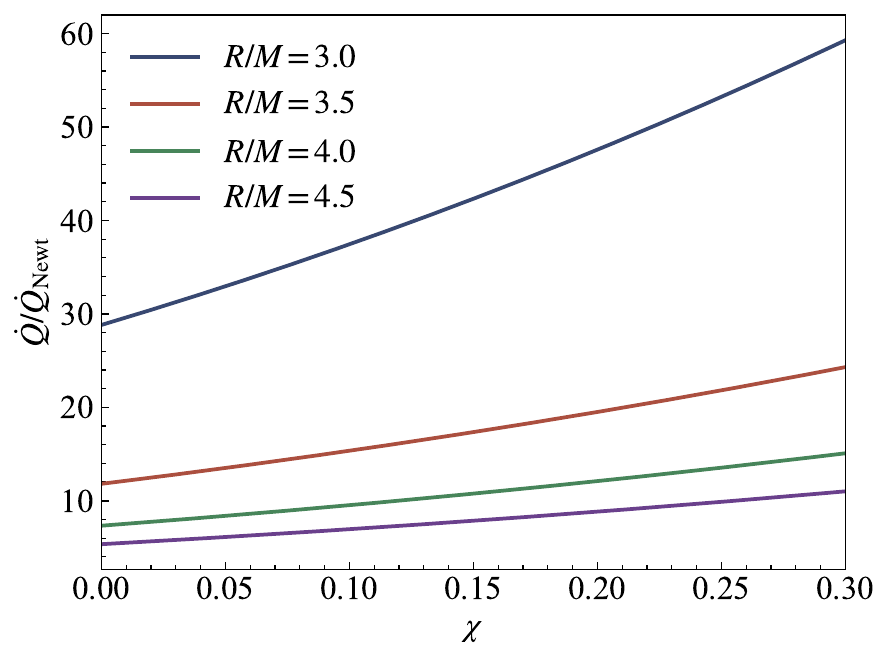}
\caption{Values of the ratio, $\dot{Q}/\dot{Q}_{Newt}$, computed at $R/M = 3,\;3.5,\;4$, and $4.5$ for several choices of the Lorentz--violating parameter $\chi$.}
\label{fig:neu_FMR_2}
\end{figure*}

\begin{table}[h!]
\centering
\caption{Emission rate, $\dot{Q}$, (in erg/s) evaluated for different choices of the Lorentz--violating parameter $\chi$ and several values of the compactness ratio $R/M$.}
\label{tab:dotQ}
\begin{tabular}{ccc}
\hline\hline
$\chi$ & $R/M$ & $\dot{Q}$ \((\text{erg/s})\) \\
\hline
0.0 & 0 & $1.50 \times 10^{50}$ \\
\hline
\multirow{2}{*}{0.0} & 3 & $4.32 \times 10^{51}$ \\
                      & 4 & $1.10 \times 10^{51}$ \\
\hline
\multirow{2}{*}{0.1} & 3 & $5.61 \times 10^{51}$ \\
                      & 4 & $1.43 \times 10^{51}$ \\
\hline
\multirow{2}{*}{0.2} & 3 & $8.89 \times 10^{51}$ \\
                      & 4 & $2.26 \times 10^{51}$ \\
\hline
\multirow{2}{*}{0.3} & 3 & $1.32 \times 10^{52}$ \\
                      & 4 & $3.35 \times 10^{51}$ \\
\hline\hline
\end{tabular}
\end{table}

The quantity $\dot{Q}$ associated with the process $\nu\bar{\nu}\to e^{+}e^{-}$ is examined in the spacetime of the new bumblebee black hole. Using the approach introduced previously, the total deposition rate is obtained by integrating the local contribution from the region exterior to the neutrino--sphere. To facilitate comparison, the results are expressed through the ratio $\dot{Q}/\dot{Q}_{Newt}$, which highlights the departure from the Newtonian prediction. The parameter $\chi$ plays a fundamental role in altering the efficiency of this energy--release mechanism. It is important to remark on the choice of the parameter range for $\chi$. Weak-field observations such as Solar System tests typically constrain Lorentz--violating parameters to magnitudes on the order of $10^{-5}$ or less. While we adhere to these stricter limits in the subsequent lensing analysis, in this section we consider larger values ranging from $0.1$ to $0.3$ for phenomenological clarity. Fig.~\ref{fig:neu_FMR_1} shows that $\dot{Q}/\dot{Q}_{Newt}$ decreases with increasing emission radius $R$, reflecting the expected weakening of the process at larger distances. Yet, for any fixed value of $R/M$, introducing a nonvanishing $\chi$ pushes the curve well above the Schwarzschild case ($\chi=0$). When the source approaches high compactness ($R/M\to 3$), the combined influence of strong redshift and trajectory modification generated by $\chi$ produces a marked enhancement in the annihilation rate.

The amplification produced by the parameter $\chi$ becomes even clearer when examining the total deposition rate displayed in Fig.~\ref{fig:neu_FMR_2}. The curves show that $\dot{Q}$ grows rapidly and nonlinearly as $\chi$ increases. Using representative luminosities and source configurations, the numerical values reported in Tab.~\ref{tab:dotQ} illustrate this trend in detail. For a highly compact emitter located at $R=3M$, the rate rises from $\dot{Q}\simeq 4.32\times10^{51}$ erg/s in the Schwarzschild case ($\chi=0$) to $\dot{Q}\simeq 8.89\times10^{51}$ erg/s when $\chi=0.3$—more than doubling the energy yield. Even for a larger emission radius, such as $R=4M$, the increase remains substantial: the rate grows from $1.10\times10^{51}$ erg/s to $2.26\times10^{51}$ erg/s for the same variation in $\chi$. In the upper range of the parameter space, e.g. $\chi=0.5$, the deposition rate reaches about $1.32\times10^{52}$ erg/s at $R=3M$. Taken together, these results indicate that the bumblebee black hole configuration can act as a markedly more effective source of neutrino--driven energy release than the Schwarzschild geometry, with $\chi$ functioning as a strong amplifier of the conversion process relevant to GRB energetics.

\begin{figure*}
\centering
\includegraphics[height=6.5cm]{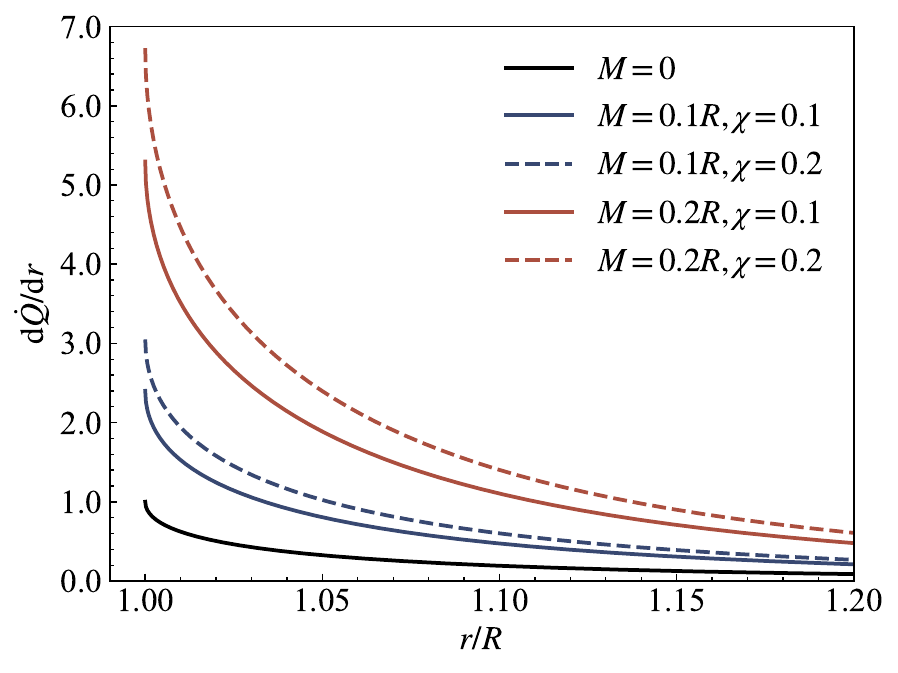}
\caption{Radial behavior of the differential deposition rate $\mathrm{d}\dot{Q}/\mathrm{d}r$ for several choices of the compactness ratio $M/R$. In the Newtonian limit ($M=0$), the expression reduces to $\mathrm{d}\dot{Q}/\mathrm{d}r = 1$ at the neutrino-sphere radius $r=R$.}
\label{fig:dQdr}
\end{figure*}

The radial behavior of the differential deposition rate $\mathrm{d}\dot{Q}/\mathrm{d}r$ is shown in Fig.~\ref{fig:dQdr}. The curves indicate that most of the annihilation takes place in a narrow region just outside the neutrino--sphere ($r/R \to 1$), consistent with expectations from standard gravity. The introduction of the parameter $\chi$ noticeably sharpens these profiles. For a fixed compactness such as $M=0.2R$ (red curves), the case $\chi=0.2$ (dashed) lies systematically above the corresponding $\chi=0.1$ result (solid) throughout the full radial domain. This behavior signals two effects: the interaction strength near the emission surface is increased, and the elevated rate persists to larger radii. Both features contribute to the substantial enhancement of the total integrated deposition rate $\dot{Q}$ in the bumblebee geometry.

A comparison with earlier analyses in bumblebee gravity helps place these results in context. Within the \textit{metric} formulation \cite{Shi:2025plr}, the Lorentz--violating parameter $\ell$ alters only the radial metric component, introducing a factor $(1+\ell)$ in $g_{rr}$ while leaving $g_{tt}$ unchanged from the Schwarzschild form. As a consequence, the resulting enhancement in the energy deposition rate is relatively mild, reaching roughly $14\%$ for $\ell=0.3$. In the \textit{metric--affine} version \cite{Shi:2025ywa}, both $g_{tt}$ and $g_{rr}$ acquire corrections, producing a stronger effect than in the purely metric case. At an emission radius of $R=4M$, the total increase amounts to about $22\%$ relative to general relativity. On the other hand, the new bumblebee black hole considered here yields a far more substantial amplification. The parameter $\chi$ modifies the temporal and radial components in a mutually reinforcing manner: $g_{tt}$ scales as $(1+\chi)^{-1}$, whereas $g_{rr}$ scales as $(1+\chi)$. Since the integrand governing the deposition rate behaves as $\sqrt{g_{rr}/(-g_{tt})^{9}}$, these combined deformations produce an effective scaling of approximately $(1+\chi)^{11/4}$. Remarkably, this high--power dependence far exceeds the enhancements obtained in either the \textit{metric} \cite{Shi:2025plr} or the \textit{metric--affine} frameworks \cite{Shi:2025ywa}. As a result, the new bumblebee geometry predicts the most efficient neutrino--driven energy release among the available bumblebee constructions, implying that, for comparable collapse conditions, it would generate the most luminous GRB counterparts.


\section{Phase Development and Oscillation Probability of Neutrinos }

The propagation of a neutrino associated with the $k$--th mass eigenstate in a spherically symmetric background can be described through a variational approach. Following the method outlined in Ref.~\cite{neu18}, and applying it to the geometry specified in Eq.~(\ref{newblacbumblebee}), one obtains the equations of motion by constructing the appropriate Lagrangian for the curved spacetime and performing the corresponding variation. The procedure proceeds as shown below:
\begin{align}
\mathcal{L}
& = -\frac{1}{2}  m_{k} g_{tt}(r,\chi)\left(\frac{\mathrm{d}t}{\mathrm{d}\Tilde{\tau}}\right)^2-\frac{1}{2}m_{k}g_{rr}(r,\chi)\left(\frac{\mathrm{d}r}{\mathrm{d}\Tilde{\tau}}\right)^2 \notag\\
& \quad -\dfrac{1}{2}m_{k}r^2\left(\frac{\mathrm{d}\theta}{\mathrm{d}\Tilde{\tau}}\right)^2  
 -\frac{1}{2}m_{k}r^2\sin^2\theta\left(\frac{\mathrm{d}\varphi}{\mathrm{d}\Tilde{\tau}}\right)^2.
\end{align}

When the motion is confined to the equatorial plane, $\theta=\pi/2$, only a restricted set of momenta contributes to the dynamics. The formulation begins by introducing the canonical momentum through
$\Tilde{p}_{\mu}=\partial \mathcal{L}/\partial(\mathrm{d}x^{\mu}/\mathrm{d}\Tilde{\tau})$, where $\mathcal{L}$ denotes the Lagrangian of the trajectory and $\Tilde{\tau}$ is the proper--time parameter. The quantity $m_{k}$ labels the mass of the $k$--th neutrino eigenmode. Under this planar symmetry, the nonvanishing components of $\Tilde{p}_{\mu}$ take simplified forms, which are listed in Refs.~\cite{neu60,Shi:2024flw}
\begin{align}
\Tilde{p}^{(k)t} &= -m_{k}g_{tt}(r,\chi)\frac{\mathrm{d}t}{\mathrm{d}\Tilde{\tau}} = -E_{k}, \\
\Tilde{p}^{(k)r} &= m_{k}g_{rr}(r,\chi)\frac{\mathrm{d}r}{\mathrm{d}\Tilde{\tau}}, \\
\Tilde{p}^{(k)\varphi} &= m_{k}r^2\frac{\mathrm{d}\varphi}{\mathrm{d}\Tilde{\tau}} = J_{k}.
\end{align}

It is important to highlight that the motion of the $k$--th neutrino eigenstate must satisfy the standard mass--shell requirement. In the present geometry, this condition takes the form
$g^{\mu\nu}(r,\chi)\,\Tilde{p}_{\mu}\Tilde{p}_{\nu}=-m_{k}^{2}$,
which acts as the fundamental constraint linking the canonical momenta to the background metric. This relation enforces consistency between the particle’s worldline and the spacetime in which it propagates \cite{neu54}
\begin{align}
-m_{k}^2 =g^{tt}(r,\chi)\Tilde{p}_t^2+g^{rr}(r,\chi)\Tilde{p}_r^2+g^{\varphi\varphi}(r,\chi)\Tilde{p}_{\varphi}^2.
\end{align}

In regions where gravitational effects can be neglected, neutrino propagation is usually described within the plane--wave scheme \cite{neu54,neu53}. Because weak interactions govern both emission and detection, the states that participate in these processes are not the mass eigenstates but linear combinations of them. These superposed configurations define the flavor eigenstates, whose properties have been extensively analyzed in earlier studies \cite{Shi:2024flw,neu62,neu63,neu61}
\ie
\ket{\nu_{\alpha}} = \sum \Tilde{U}_{\alpha i}^{*}\ket{\nu_{i}}.
\fe

A convenient description of neutrino propagation makes use of the mass eigenstates $\ket{\nu_{i}}$, each of which evolves according to its own mass and follows a distinct worldline in spacetime. Weak interactions, however, do not create or detect these mass states directly. Instead, the observable particles correspond to the flavor states $\nu_{e}$, $\nu_{\mu}$, and $\nu_{\tau}$, labeled by $\alpha=e,\mu,\tau$. The relation between the flavor basis and the mass basis is provided by a unitary $3\times3$ matrix $\Tilde{U}$, as discussed in Ref.~\cite{neu41}.

To describe the propagation process, one introduces the emission point $(t_{S},\bm{x}_{S})$ and the detection point $(t_{D},\bm{x}_{D})$. The $i$--th mass eigenstate is then evolved from the source to the detector according to
\ie
\ket{\nu_{i}(t_{D},\bm{x}_{D})}
= \exp(-\mathrm{i}\Phi_{i})\,
\ket{\nu_{i}(t_{S},\bm{x}_{S})},
\fe
where $\Phi_{i}$ denotes the phase accumulated along the corresponding spacetime trajectory. Consequently, the various mass eigenstates do not acquire the same phase during propagation. The phase associated with the $i$--th component is obtained from the relation
\ie
\Phi_{i}=\int_{\left(t_{S},\bm{x}_{S}\right)}^{\left(t_{D},\bm{x}_{D}\right)}g_{\mu\nu} (r,\chi)\Tilde{p}^{(i)\mu}\mathrm{d}x^{\nu}.
\fe

Within this approach, flavor change is analyzed by tracking how a neutrino travels from its point of emission to the location where it is detected. A state created as $\nu_{\alpha}$ may arrive as a different flavor $\nu_{\beta}$, and the likelihood of this transition is governed by the following expression:
\ie
\begin{split}
\mathcal{P}_{\alpha\beta}
& = |\left\langle \nu_{\beta}|\nu_{\alpha}\left(t_{D}, \bm{x}_{D}\right)\right\rangle|^2  = \sum_{i,j} \Tilde{U}_{\beta i} \Tilde{U}_{\beta j}^{*} \Tilde{U}_{\alpha j} \Tilde{U}_{\alpha i}^{*}\,  \exp{[-\mathrm{i}(\Phi_{i}-\Phi_{j})]}.
\end{split}
\fe

For the bumblebee black hole geometry, we restrict attention to neutrinos moving on the equatorial plane, $\theta=\pi/2$, so that their propagation is confined to a two–dimensional slice of the spacetime. Along such trajectories, each mass eigenstate acquires a phase determined by the relation shown as follows:
\ie
\begin{split}
\label{phigrandeeee}
\Phi_{k} & = \int_{\left(t_{S},\bm{x}_{S}\right)}^{\left(t_{D}, \bm{x}_{D}\right)} g_{\mu\nu} (r,\chi)\Tilde{p}^{(k)\mu}\mathrm{d}x^{\nu}\notag\\
& = \int_{\left(t_{S},\bm{x}_{S}\right)}^{\left(t_{D}, \bm{x}_{D}\right)}\left[E_{k}\mathrm{d}t - \Tilde{p}^{(k)r}\mathrm{d}r-J_{k}\mathrm{d}\varphi\right] \notag\\
& = \pm\frac{m_{k}^2}{2E_0}\int_{r_{S}}^{r_{D}}\sqrt{-g_{tt}(r,\chi)g_{rr}(r,\chi)}\left(1-\dfrac{b^2|g_{tt}(r,\chi)|}{g_{\varphi\varphi}(r,\chi)}\right)^{-\frac{1}{2}}\mathrm{d}r.
\end{split}
\fe

In regions where the gravitational influence is mild, such that $M/r \ll 1$, the integrand in Eq.~(\ref{phigrandeeee}) admits a perturbative expansion. Under this approximation, it can be written as the series is written:
\ie
\begin{split}
&\quad\sqrt{-g_{tt}(r,\chi) \,g_{rr}(r,\chi)}\left(1-\dfrac{b^2|g_{tt}(r,\chi)|}{g_{\varphi\varphi}(r,\chi)}\right)^{-1/2} \\
& \simeq\left[1-\dfrac{b^2}{r^2\left(1+\chi\right)}\right]^{-1/2}-\dfrac{b^2M}{\left(1+\chi\right)\left(r^2-\dfrac{b^2}{1+\chi}\right)^{3/2}}.
\end{split}
\fe
The corresponding phase contribution obtained from this expansion is therefore expressed as:
\ie
\Phi_k =\dfrac{m_k^2}{2E_0}\left[\sqrt{r_D^2-\dfrac{b^2}{1+\chi}}-\sqrt{r_S^2-\dfrac{b^2}{1+\chi}}+M\left(\dfrac{r_D}{\sqrt{r_D^2-\frac{b^2}{1+\chi}}}-\dfrac{r_S}{\sqrt{r_D^2-\frac{b^2}{1+\chi}}}\right)\right].
\fe

In this formulation, the relativistic neutrinos emitted by the source carry an average energy described by
$E_{0}=\sqrt{E_{k}^{2}-m_{k}^{2}}$,
where $E_{k}$ and $m_{k}$ correspond to the energy and mass of the $k$--th eigenstate. The parameter $b$, interpreted as the impact parameter, follows the definition given in Ref.~\cite{neu18}. As each neutrino travels through the curved geometry, its geodesic reaches a smallest radial value $r_{0}$ before moving outward again. When the gravitational field is weak, one may determine this turning point analytically by applying suitable approximations to the governing trajectory equation
\ie
\left(\dfrac{\mathrm{d}r}{\mathrm{d}\varphi}\right)_0=\pm\dfrac{g_{\varphi\varphi}(r,\chi)}{b^2}\sqrt{\dfrac{1}{-g_{tt}(r,\chi) \,g_{rr}(r,\chi)}-\dfrac{b^2}{g_{rr} (r,\chi)g_{\varphi\varphi}(r,\chi)}}=0.
\fe

The closest--approach radius $r_{0}$ follows from the orbital condition that governs the neutrino’s path in the weak--field limit. In this approximation, $r_{0}$ is obtained by solving the trajectory equation appropriate to that condition
\ie
\label{r0}
r_{0} \simeq \dfrac{b}{\sqrt{1+\chi}}-M.
\fe

The total phase acquired by a neutrino along its path—from the source, through the turning point $r_{0}$, and onward to the detector—follows by inserting the weak--field expression for $r_{0}$ from Eq.~(\ref{r0}) into the general phase formula of Eq.~(\ref{phigrandeeee}). Carrying out this substitution leads to the expression shown below:
\ie
\begin{split}
&\quad\Phi_{k}\left(r_{S}\to r_{0} \to r_{D}\right)\notag\\
&\simeq \frac{{m}_{k}^2}{2E_0}
\Biggl[\sqrt{r_D^2-r_0^2}+\sqrt{r_S^2-r_0^2}+M\left(\sqrt{\dfrac{r_D-r_0}{r_D+r_0}}+\sqrt{\dfrac{r_S-r_0}{r_S+r_0}}\right)\Biggr],
\end{split}
\fe
in such a way that we have
\ie
\begin{split}
\Phi_{k}
& \simeq \frac{{m}_{k}^2}{2E_0}
\Biggl\{\sqrt{r_D^2-\dfrac{b^2}{1+\chi}}+\sqrt{r_S^2-\dfrac{b^2}{1+\chi}}\notag\\
& \quad+M\Biggl[\dfrac{b}{\sqrt{\left(1+\chi\right)\,r_D^2-b^2}}+\dfrac{b}{\sqrt{\left(1+\chi\right)r_S^2-b^2}}\notag\\
&\quad+\sqrt{\dfrac{\sqrt{1+\chi}\,r_D-b}{\sqrt{1+\chi}\,r_D+b}}+\sqrt{\dfrac{\sqrt{1+\chi}\,r_S-b}{\sqrt{1+\chi}\,r_S+b}}\Biggr]\Biggr\}.
\end{split}
\fe

The next stage consists of expanding the phase expression in powers of $b/r_{S,D}$ under the assumption that the impact parameter is small compared to both the source and detector radii, $b \ll r_{S,D}$. Retaining contributions up to second order, $\mathcal{O}(b^{2}/r_{S,D}^{2})$, the phase reduces to the approximation given below:
\ie
\label{Phi_k}
\Phi_k=\dfrac{m_k^2}{2E_0}(r_D+r_S)\left[1-\dfrac{b^2}{2\left(1+\chi\right)r_Dr_S}+\dfrac{2M}{r_D+r_S}\right].
\fe

An increase in the Lorentz--violating parameter $\chi$ leads to a systematic growth in the phase accumulated by neutrinos along their trajectories. For illustration, the numerical evaluation is carried out with $E_{0}=10\,\mathrm{MeV}$, $r_{D}=10\,\mathrm{km}$, and $r_{S}=10^{5}\,r_{D}$. In this regime, the background curvature bends the neutrino paths, producing gravitational deflection. To determine the flavor--transition probability near the black hole, one must therefore account for the phase differences generated by the distinct trajectories available to the propagating neutrinos \cite{Shi:2024flw,AraujoFilho:2025rzh}
\ie
\Delta\Phi_{ij}^{pq}
= \Phi_i^{p}-\Phi_j^{q}\notag = \Delta m_{ij}^2 \mathrm{A}_{pq}+\Delta b_{pq}^2 \mathrm{B}_{ij},
\fe
in which we have
\begin{align}
\label{Delta_m}
\Delta m_{ij}^2 & = m_i^2 - m_j^2,\\
\Delta b_{pq}^2 & = b_{p}^2-b_{q}^2,\\
\mathrm{A}_{pq} & = \frac{r_{S} + r_{D}}{2 E_0}\left[1+\dfrac{2M}{r_D+r_S}-\dfrac{\sum b_{pq}^2}{4\left(1+\chi\right)r_Dr_S}\right],\\
\mathrm{B}_{ij} & = -\frac{\sum m_{ij}^2}{8E_0}\left(1+\chi\right)^{-1}\left(\frac{1}{r_{D}} + \frac{1}{r_{S}}\right),\\
\sum b_{pq}^2 & = b_{p}^2 + b_{q}^2,\\
\label{sum_m}
\sum m_{ij}^2 & = m_i^2 + m_j^2.
\end{align}

To keep track of the phases associated with the different neutrino trajectories, we label each path by a superscript and write the accumulated phase as $\Phi_{i}^{p}$, where the index $p$ identifies the route through its impact parameter $b_{p}$. The resulting phase differences that drive oscillation behavior depend on the neutrino masses $m_{i}$, the mass-squared separations $\Delta m_{ij}^{2}$, and the geometric properties of the underlying spacetime. In the limit $\chi\to 0$, the expression reduces to the familiar form obtained in Ref.~\cite{neu53}. It is also important to note that the term $\mathrm{B}_{ij}$ gathers all contributions related to the mass sector, whereas the Lorentz--violating parameter $\chi$ appears solely through its effect on the $\mathrm{A}_{pq}$ coefficient, which shapes both the size of the phase and the strength of the oscillation pattern.


\section{Deflection of Neutrino Trajectories in Curved Spacetime }

\begin{figure}
\centering
\includegraphics[height=6.5cm]{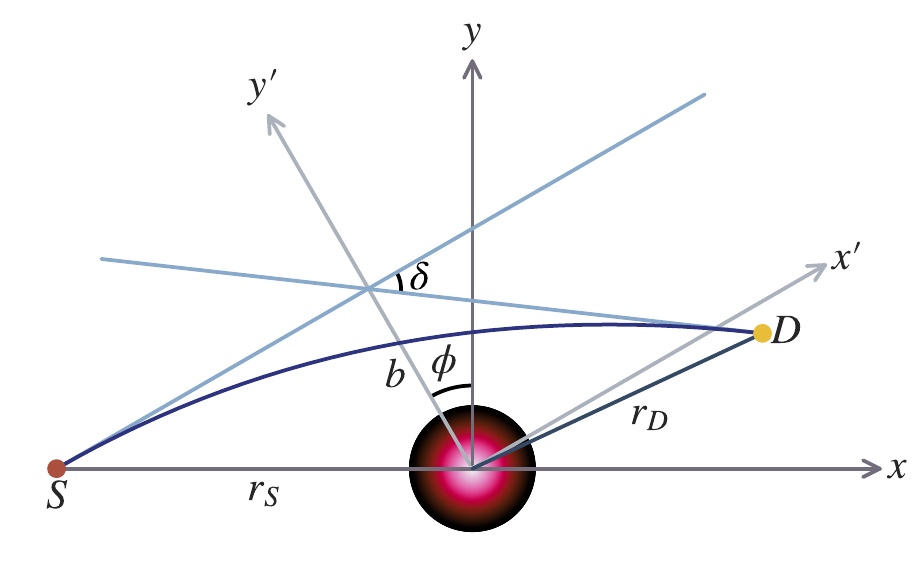}
\caption{Illustration of neutrino paths undergoing weak gravitational deflection in a curved background. The points $S$ and $D$ represent the emission site and the detector, respectively.}
\label{alaaaeasnassasiasnasag}
\end{figure}

Strong gravitational fields near compact objects can deflect neutrino paths away from radial propagation, producing lensing effects that make several trajectories link the same source to a single detector location $D$ \cite{neu54}. This situation is sketched in Fig.~\ref{alaaaeasnassasiasnasag}. When multiple routes contribute in this way, the usual single--path description of a flavor state no longer suffices; instead, the state must be represented as a coherent sum over all admissible paths \cite{neu62,neu64,neu56,neu63,Shi:2024flw,neu65,AraujoFilho:2025rzh}:
\ie
|\nu_{\alpha}(t_{D},x_{D})\rangle = N\sum_{i}\Tilde{U}_{\alpha i}^{\ast}
\sum_{p} e^{- \mathrm{i} \Phi_{i}^{p}}|\nu_{i}(t_{S}, x_{S})\rangle.
\fe

Each neutrino path is indexed by $p$, and all of these trajectories terminate at the same detector position. Because the particle can reach $D$ through several routes, the flavor conversion $\nu_{\alpha}\!\rightarrow\!\nu_{\beta}$ results from the interference among the amplitudes associated with each path. The corresponding transition probability follows from a coherent sum over all allowed trajectories and takes the form \cite{neu56,Shi:2024flw,neu64,neu62,neu63,neu65}:
\begin{align}
\label{nasndkas}
\mathcal{P}_{\alpha\beta}^{lens}  = |\langle \nu_{\beta}|\nu_{\alpha}(t_{D}, x_{D})\rangle|^{2}\notag =|N|^{2}\sum_{i, j}\Tilde{U}_{\beta i}\Tilde{U}_{\beta j}^{\ast}\Tilde{U}_{\alpha j}\Tilde{U}_{\alpha j}^{\ast}\sum_{p, q}\exp\left(\Delta\Phi_{ij}^{pq}\right).
\end{align}
Within this framework, the associated normalization factor is given by:
\begin{align}
|N|^{2} = \left[\sum_{i}|\Tilde{U}_{\alpha i}|^{2}\sum_{p,q}\exp\left(-\mathrm{i}\Delta\Phi_{ij}^{pq}\right)\right]^{-1}.
\end{align}

The phase difference $\Delta\Phi_{ij}^{pq}$ introduced earlier is the essential ingredient in determining the oscillation pattern when neutrinos propagate through a lensed geometry. As encoded in Eq.~(\ref{nasndkas}), this quantity incorporates the dependence on the neutrino masses, their squared--mass splittings, and the geometric features of the spacetime produced by the black hole. The general structure of the result resembles what has been obtained in other spherically symmetric settings, including the Schwarzschild case \cite{Shi:2024flw,neu53,AraujoFilho:2025rzh}.

The discussion now shifts to the role played by gravitational lensing in the flavor--conversion process, with particular attention to the modifications introduced by the parameter $\chi$, as we have mentioned earlier. In a weak–field lensing arrangement, and restricting the analysis to two neutrino flavors, the transition probability $\nu_{\alpha}\!\rightarrow\!\nu_{\beta}$ is evaluated by considering the geometry defined by the source, the detector, and the region where deflection occurs \cite{neu65,neu53,Shi:2024flw,neu54}
\ie
\begin{split}
\label{fssadasdasdasd}
\mathcal{P}_{\alpha\beta}^{lens}
&=\left|N\right|^2\biggl\{2\sum_i\left|\Tilde{U}_{\beta i}\right|^2\left|\Tilde{U}_{\alpha i}\right|^2\left[1+\cos\left(\Delta b_{12}^2\mathrm{B}_{ii}\right)\right]\notag\\
&\quad+\sum_{i\neq j}\Tilde{U}_{\beta i}\Tilde{U}_{\beta j}^{*}\Tilde{U}_{\alpha j}\Tilde{U}_{\alpha i}^*\left[\exp\left(-\mathrm{i}\Delta m_{ij}^2 \mathrm{A}_{11}\right)+\exp\left(-\mathrm{i}\Delta m_{ij}^2 \mathrm{A}_{22}\right)\right]\notag\\
&\quad+\sum_{i\neq j}\Tilde{U}_{\beta i}\Tilde{U}_{\beta j}^{*}\Tilde{U}_{\alpha j}\Tilde{U}_{\alpha i}^*\exp\left(-\mathrm{i}\Delta b_{12}^2\mathrm{B}_{ij}\right)\exp\left(-\mathrm{i}\Delta m_{ij}^2\mathrm{A}_{12}\right)\notag\\
&\quad+\sum_{i\neq j}\Tilde{U}_{\beta i}\Tilde{U}_{\beta j}^{*}\Tilde{U}_{\alpha j}\Tilde{U}_{\alpha i}^{*}\exp\left(\mathrm{i}\Delta b_{21}^2\mathrm{B}_{ij}\right)\exp\left(-\mathrm{i}\Delta m_{ij}^2\mathrm{A}_{21}\right)\biggr\}.
\end{split}
\fe

Equation~(\ref{fssadasdasdasd}) organizes the flavor–transition probability into a collection of contributions, each associated with a particular combination of mass indices and propagation paths. Terms with identical mass labels ($i=j$) describe the isolated evolution of a single mass eigenstate and therefore contain no interference effects. When the mass indices differ but the neutrino follows the same geodesic ($i\neq j$, $p=q$), the contribution arises from the phase mismatch accumulated by different eigenstates along one and the same path.

Additional interference appears when both the mass labels and the trajectory indices are distinct ($i\neq j$, $p\neq q$). These mixed contributions are evaluated separately for $p<q$ and $p>q$, since the differing trajectory lengths and curvature effects generate asymmetric phase accumulations along each route. If the analysis is restricted to two flavor species, the overall structure simplifies considerably. In this case, the flavor and mass bases are connected through a $2\times2$ unitary matrix characterized solely by the mixing angle $\alpha$ \cite{neu43}
\ie
\label{U}
\Tilde{U}\equiv\left(\begin{matrix}
\cos\alpha&\sin\alpha\\
-\sin\alpha&\cos\alpha
\end{matrix}\right).
\fe

By inserting the mixing matrix of Eq.~(\ref{U}) into the general probability formula of Eq.~(\ref{fssadasdasdasd}), one obtains the explicit expression governing the transition
$\nu_{e}\!\rightarrow\!\nu_{\mu}$
\ie
\begin{split}
\label{aparaoabaalall}
\mathcal{P}_{\alpha\beta}^{lens}
&=\left|N\right|^2\sin^{2}2\alpha\biggl[\sin^2\left(\dfrac{1}{2}\Delta m_{12}^2\mathrm{A}_{11}\right)+\sin^2\left(\dfrac{1}{2}\Delta m_{12}^2\mathrm{A}_{22}\right)\notag\\
&\quad+\dfrac{1}{2}\cos\left(\Delta b_{12}^2\mathrm{B}_{11}\right)+\dfrac{1}{2}\cos\left(\Delta b_{12}^2\mathrm{B}_{22}\right)-\cos\left(\Delta b_{12}^2\mathrm{B}_{12}\right)\cos\left(\Delta m_{12}^2\mathrm{A}_{12}\right)\biggr].
\end{split}
\fe

Using the mixing matrix of Eq.~(\ref{U}) together with the phase differences associated with the various neutrino paths, the corresponding normalization factor is obtained in the form shown below:
\ie
\left|N\right|^2 =\left[2+2\cos^2\alpha\cos\left(\Delta b_{12}^2\mathrm{B}_{11}\right)+2\sin^2\alpha\cos\left(\Delta b_{12}^2\mathrm{B}_{22}\right)\right]^{-1}.
\fe


\section{Computational Analysis }

To investigate how oscillations behave in the curved geometry produced by the black hole, the lensing probabilities given in Eq.~(\ref{aparaoabaalall}) must be evaluated. In the $(x,y)$ coordinate system, the lens is positioned at the origin, while the source and detector lie at the radial locations $r_{S}$ and $r_{D}$. For practical calculations, it is convenient to introduce a rotated frame $(x',y')$ obtained by turning the original coordinates by an angle $\varphi$. This transformation is expressed as \cite{neu53,Shi:2024flw}:
\ie
x' = x\cos\varphi + y\sin\varphi, \quad y' = -x\sin\varphi + y\cos\varphi.
\nonumber
\fe

Choosing $\varphi=0$ corresponds to a particularly simple geometric setup in which the source, lens, and detector all lie on the same straight line in the plane. With this alignment, the three points share a common axis, allowing the propagation analysis to proceed with a reduced set of geometric ingredients.

In the weak-lensing regime, the bending of a neutrino trajectory—described by the deflection angle $\delta$—is determined by the value of the impact parameter $b$, as shown in Refs.~\cite{neu53,Shi:2024flw}. Neutrinos emitted at $S$ pass near the bumblebee black hole, experience gravitational deflection, and eventually reach the detector at $D$. In the $(x,y)$ coordinates, the distances from the source to the lens and from the lens to the detector are represented by $r_{S}$ and $r_{D}$.

A detailed derivation of the deflection equation for the bumblebee black hole was provided in Ref.~\cite{AraujoFilho:2025zaj}, where the lensing relation was obtained directly from the geodesic equations of the modified geometry. The result takes the form:
\ie
\label{daealataa}
\delta \sim\frac{y_{D}'- b}{x_{D}'}=\sqrt{1+\chi}\left(\pi+\dfrac{4M}{b}\right)-\pi.
\fe

Placing the detector at the point $(x_{D}',y_{D}')$ in the rotated frame and using the relation $\sin\varphi=b/r_{S}$ allows the deflection formula of Eq.~(\ref{daealataa}) to be recast in the following form:
\ie
\begin{split}
\label{solve_b}
&\quad\left[by_D-\left(1-\sqrt{1+\chi}\right)\pi bx_D+4M\sqrt{1+\chi}x_D\right]\sqrt{1-\dfrac{b^2}{r_S^2}}\notag\\
&=b^2\left[\dfrac{x_D}{r_S}+1+\left(1-\sqrt{1+\chi}\right)\pi\dfrac{y_D}{r_S}\right]-\sqrt{1+\chi}\dfrac{4Mby_D}{r_S}.
\end{split}
\fe

\begin{figure*}
\centering
\includegraphics[height=20.5cm]{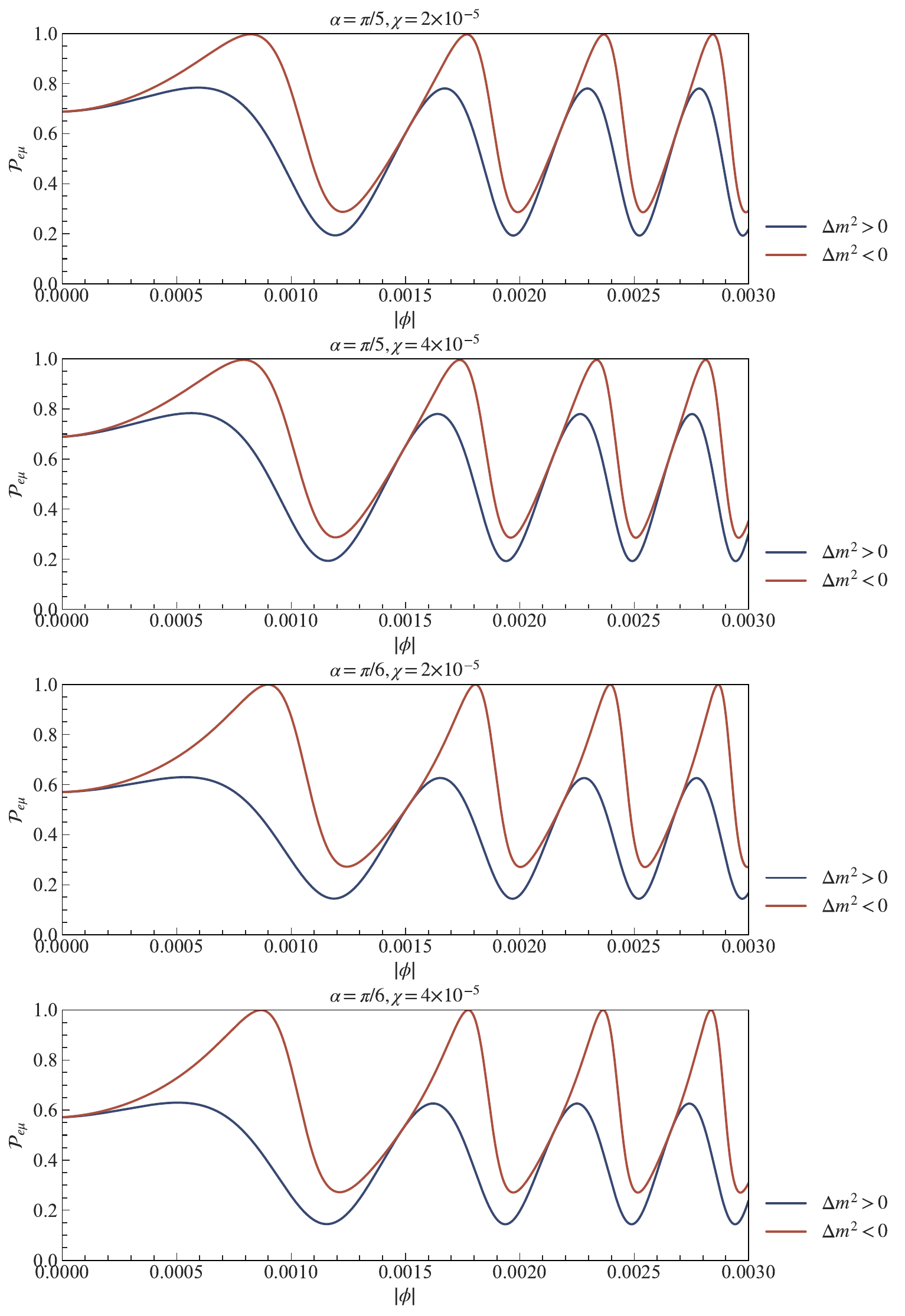}
\caption{\label{fig:prob1} Transition probability $\nu_{e}\!\rightarrow\nu_{\mu}$ as a function of the azimuthal angle $\varphi$ for $\chi = 2,4 \times 10^{-5}$, evaluated at fixed mixing angles $\alpha = \pi/5$ and $\pi/6$.}
\end{figure*}

\begin{figure*}
\centering
\includegraphics[height=20.5cm]{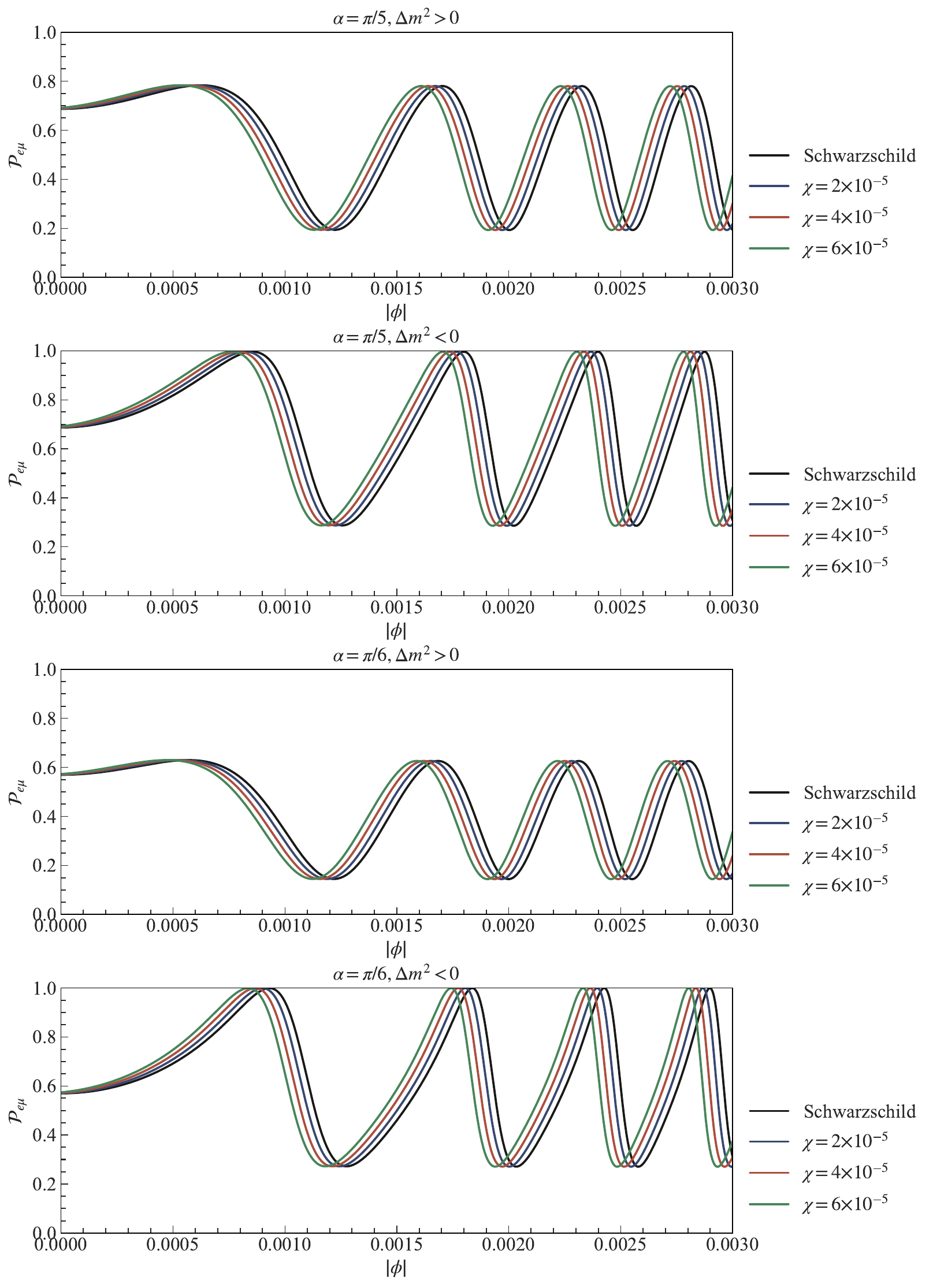}
\caption{\label{fig:prob2} Oscillation probability for the transition $\nu_{e}\!\rightarrow\nu_{\mu}$ plotted against the azimuthal angle $\varphi$ in a two--flavor setup, shown for both normal and inverted hierarchies. The curves correspond to $\chi = 2,4,6 \times 10^{-5}$ and mixing angles $\alpha = \pi/5$ and $\pi/6$.}
\end{figure*}

\begin{figure*}
\centering
\includegraphics[height=20.5cm]{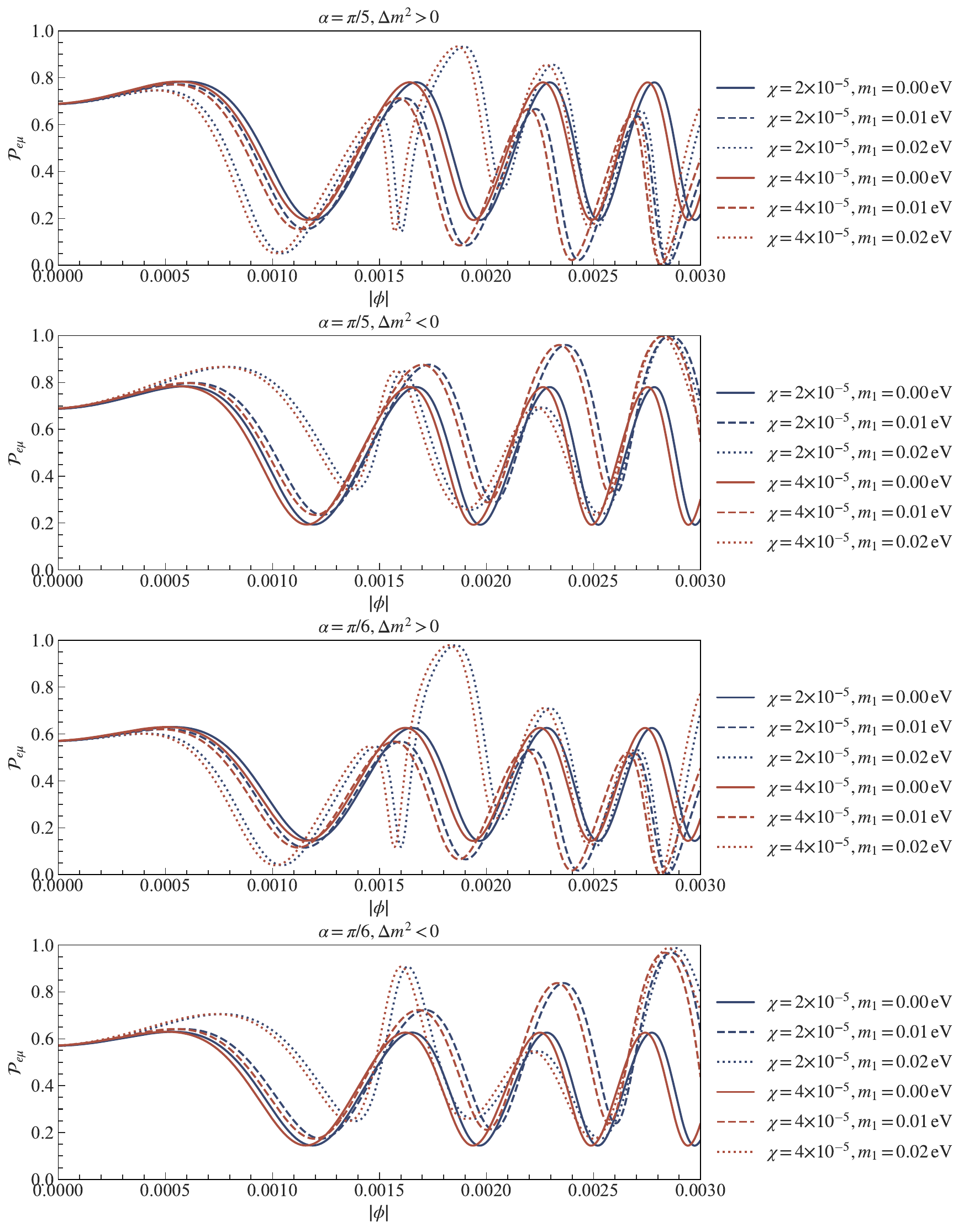}
\caption{\label{fig:prob3} Dependence of the $\nu_{e}\!\rightarrow\nu_{\mu}$ transition probability on the azimuthal angle $\varphi$ for both normal ($\Delta m^{2}>0$) and inverted ($\Delta m^{2}<0$) hierarchies. The blue and red curves correspond to $\chi = 2\times10^{-10}$ and $4\times10^{-10}$, respectively.}
\end{figure*}

In this section, neutrino flavor conversion is examined in the spacetime generated by the bumblebee black hole, where gravitational effects enter through the Lorentz--violating parameter $\chi$. The detector is assumed to follow a circular orbit around the Sun, parametrized by $x_{D}=r_{D}\cos\varphi$ and $y_{D}=r_{D}\sin\varphi$. Within this geometric setup, the quartic equation given in Eq.~\eqref{solve_b} can be solved numerically in the equatorial plane, producing two positive real impact parameters, $b_{1}$ and $b_{2}$, for each value of the azimuthal angle $\varphi$. The individual components of this polynomial follow from Eqs.~\eqref{Delta_m}–\eqref{sum_m}, and the evaluation proceeds by inserting the chosen values of $m_{1}$, $m_{2}$, $b_{1}$, and $b_{2}$.

The resulting oscillation probabilities were contrasted between the bumblebee geometry ($\chi\neq 0$) and the Schwarzschild limit ($\chi=0$). Figures~\ref{fig:prob1}, \ref{fig:prob2}, and \ref{fig:prob3} display how variations in the azimuthal angle $\varphi$ and in the mass--squared difference $\Delta m^{2}$ shape the behavior of $\nu_{e}\!\rightarrow\nu_{\mu}$ transitions.

We now examine quantum–interference effects experienced by neutrinos as they travel through the gravitational environment of the new bumblebee black hole. Working within a two--flavor scheme, the transition probability $\mathcal{P}_{e\mu}$ is evaluated in the presence of gravitational lensing. The goal is to clarify how the Lorentz--violating parameter $\chi$, the choice of mass ordering, and the absolute neutrino mass scale jointly influence the resulting oscillation behavior.

We begin by assessing the role of the mass ordering, as shown in Fig.~\ref{fig:prob1}. Across all cases, a clear pattern emerges: the Inverted Ordering ($\Delta m^{2}<0$, red curves) systematically produces larger transition probabilities than the Normal Ordering ($\Delta m^{2}>0$, blue curves). This distinction originates from how the vacuum phase combines with the phase generated by the curved geometry. For the Inverted Ordering, the negative value of $\Delta m^{2}_{31}$ shifts the total phase in a way that enhances the contribution coming from the bumblebee background. In contrast, when the ordering is normal, the positive sign of $\Delta m^{2}_{31}$ leads to a partial reduction of the geometric contribution, yielding a smaller overall probability. In effect, the geometry acts as a selective amplifier: it strengthens the signal associated with the Inverted Ordering while diminishing that of the normal one.

Fig.~\ref{fig:prob2} shows how the parameter $\chi$ alters the oscillation phase. As $\chi$ increases from $0$ to $6\times10^{-5}$, the oscillation pattern shifts noticeably. This strong sensitivity originates from the way the modified geometry changes the effective optical distance traveled by the neutrino. In the weak--field regime, the phase satisfies $\Phi \propto \int \sqrt{g_{rr}/(-g_{tt})}\,\mathrm{d}r$. For our new bumblebee solution, the metric behaves as $g_{tt}\sim (1+\chi)^{-1}$ and $g_{rr}\sim (1+\chi)$, which introduces an overall factor of $(1+\chi)$ inside the square root. In contrast to other alternative–gravity scenarios—where metric corrections may partially cancel or appear only at higher order—the parameter $\chi$ enters the phase integrand in a direct and linear manner. As a result, even values as small as $\chi\sim10^{-5}$ generate a sizable cumulative phase shift over large propagation distances. This explains the rapid displacement of the oscillation fringes displayed in the figure.

The influence of the absolute mass scale $m_{1}$ is displayed in Fig.~\ref{fig:prob3}. The increasingly intricate pattern that appears when $m_{1}$ is raised from $0\,\mathrm{eV}$ to $0.02\,\mathrm{eV}$ can be attributed to a dephasing mechanism. The total phase includes a kinetic contribution of the form $m_{i}^{2}/(2E)$; when $m_{1}\approx 0$, the behavior is governed primarily by the mass splitting $\Delta m^{2}$. As $m_{1}$ grows, however, the individual masses begin to provide a non-negligible contribution to the phase, becoming comparable to the geometric correction induced by the spacetime.  In the Inverted Ordering, the additional kinetic term counteracts the geometric enhancement discussed earlier, producing a reduction in the peak transition probabilities over certain angular intervals.

To place these results in a broader perspective, we contrast the behavior of the new bumblebee geometry with earlier Lorentz--violating constructions, namely the \textit{metric} and \textit{metric--affine} formulations \cite{Shi:2025plr,Shi:2025ywa}. In the \textit{metric} approach, the deformation enters exclusively through the radial component, $g_{rr}\rightarrow (1+\ell)g_{rr}$, while $g_{tt}$ remains unchanged from its Schwarzschild value. The resulting phase modification is therefore mild, introducing a factor close to $\sqrt{1+\ell}\simeq 1+\ell/2$ for small $\ell$. The \textit{metric-affine} realization alters both metric components, encoded in the functions $X_{1}$ and $X_{2}$ that arise from the underlying non-metricity \cite{Shi:2025ywa}. Although the simultaneous modification of $g_{tt}$ and $g_{rr}$ amplifies the net effect compared with the purely metric scenario, the correction typically appears at perturbative order, scaling as $1+\mathcal{O}(X)$.
The situation is markedly different for the new bumblebee black hole. Here, the time and radial components are deformed by reciprocal factors, $g_{tt}\sim (1+\chi)^{-1}$ and $g_{rr}\sim (1+\chi)$, which combine to rescale the optical metric by an entire factor of $(1+\chi)$. For small parameters, this yields a significantly stronger response—roughly twice the sensitivity seen in the \textit{metric} case, where the correction grows only as $1+\ell/2$ \cite{Shi:2025plr}.

This comparison has direct implications for observational tests. Because the new bumblebee geometry produces a substantially stronger phase response, experiments with the sensitivity of IceCube-Gen2 or KM3NeT would be able to explore the parameter $\chi$ far more effectively than the deformation parameters appearing in earlier models. In the absence of measurable deviations from standard oscillation behavior, the resulting limits on $\chi$ would be roughly twice as restrictive as those derived for $\ell$ in the metric formulation. On the other hand, should a deviation be observed, the present framework naturally generates sizable effects with small and physically reasonable values of $\chi$, eliminating the need for any fine adjustments of the parameter.


\section{Conclusion}\label{Sec:Conclusion}

The analysis carried out in this work showed how the black hole geometry emerging from spontaneous Lorentz--symmetry breaking in bumblebee gravity reshaped several aspects of neutrino propagation. The parameter $\chi$, which controlled the deformation of both temporal and radial components of the metric, permeated all three sectors examined here: the deposition of energy through neutrino--antineutrino annihilation, the phase acquired by propagating mass eigenstates, and the lensing--driven interference pattern governing flavor conversion.

In the first part of the study, the modified geometry substantially increased the efficiency of the $\nu\bar{\nu}\rightarrow e^{+}e^{-}$ channel. Because the deformation rescaled the optical metric through the combination $g_{tt}\sim (1+\chi)^{-1}$ and $g_{rr}\sim (1+\chi)$, the resulting contribution to $\dot{Q}$ grew rapidly with $\chi$. The enhancement was significantly stronger than those found in earlier \textit{metric} and \textit{metric--affine} versions of bumblebee gravity, placing this new solution as the most effective scenario for neutrino--powered energy release among the available Lorentz--violating configurations.

The second component of the investigation focused on the evolution of the oscillation phase. The background curvature introduced a correction proportional to the integrated factor $\sqrt{g_{rr}/(-g_{tt})}$, which received a linear contribution from $\chi$. Even small deformations generated a noticeable shift in the phase accumulated between the source and the detector, affecting the interference responsible for flavor transitions. The effect was absent in the standard Schwarzschild description and emerged distinctly when compared with the milder modifications present in earlier bumblebee models.

The third part addressed how weak gravitational lensing altered oscillation behavior once neutrinos reached the detector through multiple trajectories. The phase differences associated with each path were highly sensitive to the geometry, and the deformation parameter $\chi$ played a decisive role in reshaping the angular dependence of the transition probability. Differences between the normal and inverted mass hierarchies were amplified, and the absolute mass scale $m_{1}$ also introduced additional structure in the resulting oscillation patterns.

Taken together, these results indicated that the new bumblebee black hole provided a setting in which Lorentz--violating effects left a clear mark on neutrino propagation. The combined enhancement of the annihilation channel, the sensitivity of the accumulated phase, and the modifications induced by lensing suggested that $\chi$ could be probed effectively through astrophysical neutrino observations. Future detectors with improved angular and flavor resolution would therefore be capable of imposing stringent limits on this deformation parameter or might be revealing deviations associated with spontaneous Lorentz--symmetry breaking.

As a further perspective, it would be worthwhile to examine the same set of phenomena discussed here in the context of the Kalb--Ramond black hole introduced in Ref.~\cite{Liu:2024oas}. This analysis is currently in its final stages of preparation and will be made available on arXiv very soon.


\section*{Acknowledgments}
\hspace{0.5cm} A.A.A.F. is supported by Conselho Nacional de Desenvolvimento Cient\'{\i}fico e Tecnol\'{o}gico (CNPq) and Fundação de Apoio à Pesquisa do Estado da Paraíba (FAPESQ), project numbers 150223/2025-0 and 1951/2025. 

\section*{Data Availability Statement}

Data Availability Statement: No Data associated with the manuscript

\bibliographystyle{ieeetr}
\bibliography{main}

\begin{thebibliography}{10}

\bibitem{kostelecky1989spontaneous}
V.~A. Kosteleck{\`y} and S.~Samuel, ``Spontaneous breaking of lorentz symmetry
  in string theory,'' {\em Physical Review D}, vol.~39, no.~2, p.~683, 1989.

\bibitem{colladay1997cpt}
D.~Colladay and V.~A. Kosteleck{\`y}, ``Cpt violation and the standard model,''
  {\em Physical Review D}, vol.~55, no.~11, p.~6760, 1997.

\bibitem{kostelecky2004gravity}
V.~A. Kosteleck{\`y}, ``Gravity, lorentz violation, and the standard model,''
  {\em Physical Review D}, vol.~69, no.~10, p.~105009, 2004.

\bibitem{kostelecky1999constraints}
V.~A. Kosteleck{\`y} and C.~D. Lane, ``Constraints on lorentz violation from
  clock-comparison experiments,'' {\em Physical Review D}, vol.~60, no.~11,
  p.~116010, 1999.

\bibitem{kostelecky2011data}
V.~A. Kosteleck{\`y} and N.~Russell, ``Data tables for lorentz and cpt
  violation,'' {\em Reviews of Modern Physics}, vol.~83, no.~1, pp.~11--31,
  2011.

\bibitem{Bluhm:2023kph}
R.~Bluhm and Y.~Zhi, ``{Spontaneous and Explicit Spacetime Symmetry Breaking in
  Einstein{\textendash}Cartan Theory with Background Fields},'' {\em Symmetry},
  vol.~16, no.~1, p.~25, 2024.

\bibitem{bluhm2005spontaneous}
R.~Bluhm and V.~A. Kosteleck{\`y}, ``Spontaneous lorentz violation,
  nambu-goldstone modes, and gravity,'' {\em Physical Review D—Particles,
  Fields, Gravitation, and Cosmology}, vol.~71, no.~6, p.~065008, 2005.

\bibitem{Bluhm:2019ato}
R.~Bluhm, H.~Bossi, and Y.~Wen, ``{Gravity with explicit spacetime symmetry
  breaking and the Standard-Model Extension},'' {\em Phys. Rev. D}, vol.~100,
  no.~8, p.~084022, 2019.

\bibitem{Maluf:2014dpa}
R.~V. Maluf, C.~A.~S. Almeida, R.~Casana, and M.~M. Ferreira, Jr.,
  ``{Einstein-Hilbert graviton modes modified by the Lorentz-violating
  bumblebee Field},'' {\em Phys. Rev. D}, vol.~90, no.~2, p.~025007, 2014.

\bibitem{bluhm2008spontaneous}
R.~Bluhm, S.-H. Fung, and V.~A. Kosteleck{\`y}, ``Spontaneous lorentz and
  diffeomorphism violation, massive modes, and gravity,'' {\em Physical Review
  D—Particles, Fields, Gravitation, and Cosmology}, vol.~77, no.~6,
  p.~065020, 2008.

\bibitem{Maluf:2013nva}
R.~V. Maluf, V.~Santos, W.~T. Cruz, and C.~A.~S. Almeida, ``{Matter-gravity
  scattering in the presence of spontaneous Lorentz violation},'' {\em Phys.
  Rev. D}, vol.~88, no.~2, p.~025005, 2013.

\bibitem{kostelecky1991photon}
V.~A. Kosteleck{\`y} and S.~Samuel, ``Photon and graviton masses in string
  theories,'' {\em Physical Review Letters}, vol.~66, no.~14, p.~1811, 1991.

\bibitem{jacobson2004einstein}
T.~Jacobson and D.~Mattingly, ``Einstein-aether waves,'' {\em Physical Review
  D}, vol.~70, no.~2, p.~024003, 2004.

\bibitem{Bertolami:2005bh}
O.~Bertolami and J.~Paramos, ``{The Flight of the bumblebee: Vacuum solutions
  of a gravity model with vector-induced spontaneous Lorentz symmetry
  breaking},'' {\em Phys. Rev. D}, vol.~72, p.~044001, 2005.

\bibitem{Casana:2017jkc}
R.~Casana, A.~Cavalcante, F.~P. Poulis, and E.~B. Santos, ``{Exact
  Schwarzschild-like solution in a bumblebee gravity model},'' {\em Phys. Rev.
  D}, vol.~97, no.~10, p.~104001, 2018.

\bibitem{Liu:2024wpa}
W.~Liu, C.~Wen, and J.~Wang, ``{Lorentz violation alleviates gravitationally
  induced entanglement degradation},'' {\em JHEP}, vol.~01, p.~184, 2025.

\bibitem{AraujoFilho:2025hkm}
A.~A. Ara{\'u}jo~Filho, ``{How does non-metricity affect particle creation and
  evaporation in bumblebee gravity?},'' {\em JCAP}, vol.~06, p.~026, 2025.

\bibitem{AraujoFilho:2024ctw}
A.~A. Ara{\'u}jo~Filho, ``{Particle creation and evaporation in Kalb-Ramond
  gravity},'' {\em JCAP}, vol.~04, p.~076, 2025.

\bibitem{Neves:2022qyb}
J.~C.~S. Neves, ``{Kasner cosmology in bumblebee gravity},'' {\em Annals
  Phys.}, vol.~454, p.~169338, 2023.

\bibitem{Neves:2024ggn}
J.~C.~S. Neves and F.~G. Gardim, ``{Stars and quark stars in bumblebee
  gravity},'' {\em Annals Phys.}, vol.~475, p.~169950, 2025.

\bibitem{Liang:2022hxd}
D.~Liang, R.~Xu, X.~Lu, and L.~Shao, ``{Polarizations of gravitational waves in
  the bumblebee gravity model},'' {\em Phys. Rev. D}, vol.~106, no.~12,
  p.~124019, 2022.

\bibitem{amarilo2024gravitational}
K.~M. Amarilo, M.~B. Ferreira~Filho, A.~A. Ara{\'u}jo~Filho, and J.~A. A.~S.
  Reis, ``Gravitational waves effects in a lorentz--violating scenario,'' {\em
  Physics Letters B}, vol.~855, p.~138785, 2024.

\bibitem{Maluf:2020kgf}
R.~V. Maluf and J.~C.~S. Neves, ``{Black holes with a cosmological constant in
  bumblebee gravity},'' {\em Phys. Rev. D}, vol.~103, no.~4, p.~044002, 2021.

\bibitem{Kumar:2025bim}
A.~Kumar, S.~U. Islam, and S.~G. Ghosh, ``{Probing Lorentz Symmetry Violation
  through Lensing Observables of Rotating Black Holes},'' 8 2025.

\bibitem{Filho:2022yrk}
A.~A.~A. Filho, J.~R. Nascimento, A.~Y. Petrov, and P.~J. Porf{\'\i}rio,
  ``{Vacuum solution within a metric-affine bumblebee gravity},'' {\em Phys.
  Rev. D}, vol.~108, no.~8, p.~085010, 2023.

\bibitem{AraujoFilho:2024ykw}
A.~A. Ara{\'u}jo~Filho, J.~R. Nascimento, A.~Y. Petrov, and P.~J.
  Porf{\'\i}rio, ``{An exact stationary axisymmetric vacuum solution within a
  metric-affine bumblebee gravity},'' {\em JCAP}, vol.~07, p.~004, 2024.

\bibitem{AraujoFilho:2025rvn}
A.~A. Ara{\'u}jo~Filho, N.~Heidari, I.~P. Lobo, Y.~Shi, and F.~S.~N. Lobo,
  ``{The Flight of the Bumblebee in a Non-Commutative Geometry: A New Black
  Hole Solution},'' 9 2025.

\bibitem{AraujoFilho:2025jcu}
A.~A. Ara{\'u}jo~Filho, N.~Heidari, and I.~P. Lobo, ``{A non-commutative
  Kalb-Ramond black hole},'' {\em JCAP}, vol.~09, p.~076, 2025.

\bibitem{Magalhaes:2025nql}
R.~B. Magalh{\~a}es, L.~A. Lessa, and M.~M. Ferreira, ``{Wormholes in
  Lorentz-violating gravity},'' 5 2025.

\bibitem{AraujoFilho:2024iox}
A.~A. Ara{\'u}jo~Filho, J.~A. A.~S. Reis, and A.~{\"O}vg{\"u}n, ``{Modified
  particle dynamics and thermodynamics in a traversable wormhole in bumblebee
  gravity},'' {\em Eur. Phys. J. C}, vol.~85, no.~1, p.~83, 2025.

\bibitem{Magalhaes:2025lti}
R.~B. Magalh{\~a}es, L.~A. Lessa, and R.~Casana, ``{Lorentz-violating
  wormholes: The role of the matter coupled to Lorentz-violating fields},'' 7
  2025.

\bibitem{Ovgun:2018xys}
A.~{\"O}vg{\"u}n, K.~Jusufi, and {\.I}.~Sakall{\i}, ``{Exact traversable
  wormhole solution in bumblebee gravity},'' {\em Phys. Rev. D}, vol.~99,
  no.~2, p.~024042, 2019.

\bibitem{Pereira:2025xnw}
C.~F.~S. Pereira, M.~V. d.~S. Silva, H.~Belich, D.~C.~Rodrigues, J.~C. Fabris,
  and M.~E. Rodrigues, ``{Black-bounce solutions in a k-essence theory under
  the effects of bumblebee gravity},'' {\em Phys. Rev. D}, vol.~111, no.~12,
  p.~124005, 2025.

\bibitem{Shi:2025plr}
Y.~Shi and A.~A. Ara{\'u}jo~Filho, ``Effects of bumblebee gravity on neutrino
  motion,'' {\em Journal of Cosmology and Astroparticle Physics}, vol.~2025,
  no.~11, p.~045, 2025.

\bibitem{Shi:2025ywa}
Y.~Shi and A.~A. Ara{\'u}jo~Filho, ``The role of non-metricity on neutrino
  behavior in bumblebee gravity,'' {\em arXiv preprint arXiv:2505.12551}, 2025.

\bibitem{Shi:2025rfq}
Y.~Shi and A.~A. Ara{\'u}jo~Filho, ``{Influence of a Kalb-Ramond black hole on
  neutrino behavior},'' {\em JHEP}, vol.~08, p.~028, 2025.

\bibitem{Pontecorvo2}
B.~Pontecorvo, ``Neutrino experiments and the problem of conservation of
  leptonic charge,'' {\em Sov. Phys. JETP}, vol.~26, no.~984-988, p.~165, 1968.

\bibitem{Pontecorvo1}
B.~Pontecorvo, ``Mesonium and antimesonium,'' {\em Soviet Journal of
  Experimental and Theoretical Physics}, vol.~6, p.~429, 1958.

\bibitem{maki1962remarks}
Z.~Maki, M.~Nakagawa, and S.~Sakata, ``Remarks on the unified model of
  elementary particles,'' {\em Progress of theoretical physics}, vol.~28,
  no.~5, pp.~870--880, 1962.

\bibitem{neu44}
F.~An, J.~Bai, A.~Balantekin, H.~Band, D.~Beavis, W.~Beriguete, M.~Bishai,
  S.~Blyth, K.~Boddy, R.~Brown, {\em et~al.}, ``Observation of
  electron-antineutrino disappearance at daya bay,'' {\em Physical Review
  Letters}, vol.~108, no.~17, p.~171803, 2012.

\bibitem{neu42}
P.~D. Group, P.~Zyla, R.~Barnett, J.~Beringer, O.~Dahl, D.~Dwyer, D.~Groom,
  C.-J. Lin, K.~Lugovsky, E.~Pianori, {\em et~al.}, ``Review of particle
  physics,'' {\em Progress of Theoretical and Experimental Physics}, vol.~2020,
  no.~8, p.~083C01, 2020.

\bibitem{neu43}
I.~Esteban, M.~C. Gonz{\'a}lez-Garc{\'\i}a, A.~Hernandez-Cabezudo, M.~Maltoni,
  and T.~Schwetz, ``Global analysis of three-flavour neutrino oscillations:
  synergies and tensions in the determination of $\theta$23, $\delta$cp, and
  the mass ordering,'' {\em Journal of High Energy Physics}, vol.~2019, no.~1,
  pp.~1--35, 2019.

\bibitem{neu40}
P.~F. de~Salas, D.~V. Forero, C.~A. Ternes, M.~Tortola, and J.~W.~F. Valle,
  ``{Status of neutrino oscillations 2018: 3$\sigma$ hint for normal mass
  ordering and improved CP sensitivity},'' {\em Phys. Lett. B}, vol.~782,
  pp.~633--640, 2018.

\bibitem{neu41}
I.~Esteban, M.~C. Gonz{\'a}lez-Garc{\'\i}a, A.~Hernandez-Cabezudo, M.~Maltoni,
  and T.~Schwetz, ``Global analysis of three-flavour neutrino oscillations:
  synergies and tensions in the determination of $\theta$23, $\delta$cp, and
  the mass ordering,'' {\em Journal of High Energy Physics}, vol.~2019, no.~1,
  pp.~1--35, 2019.

\bibitem{neu39}
F.~Capozzi, G.~L. Fogli, E.~Lisi, A.~Marrone, D.~Montanino, and A.~Palazzo,
  ``{Status of three-neutrino oscillation parameters, circa 2013},'' {\em Phys.
  Rev. D}, vol.~89, p.~093018, 2014.

\bibitem{neu45}
P.~D. Group {\em et~al.}, ``Review of particle physics,'' {\em Physical Review
  D}, vol.~98, no.~3, p.~030001, 2018.

\bibitem{Chakrabarty:2023kld}
H.~Chakrabarty, A.~Chatrabhuti, D.~Malafarina, B.~Silasan, and T.~Tangphati,
  ``{Effects of gravitational lensing by Kaluza-Klein black holes on neutrino
  oscillations},'' {\em JCAP}, vol.~08, p.~018, 2023.

\bibitem{Shi:2024flw}
Y.~Shi and H.~Cheng, ``{The neutrino flavor oscillations in the static and
  spherically symmetric black-hole-like wormholes},'' {\em Eur. Phys. J. C},
  vol.~85, no.~8, p.~909, 2025.

\bibitem{AraujoFilho:2025rzh}
A.~A. Ara{\'u}jo~Filho, N.~Heidari, and Y.~Shi, ``Neutrino dynamics in a
  non-commutative spacetime,'' {\em arXiv e-prints}, pp.~arXiv--2504, 2025.

\bibitem{Alloqulov:2024sns}
M.~Alloqulov, H.~Chakrabarty, D.~Malafarina, B.~Ahmedov, and A.~Abdujabbarov,
  ``{Gravitational lensing of neutrinos in parametrized black hole
  spacetimes},'' {\em JCAP}, vol.~02, p.~070, 2025.

\bibitem{Shi:2023kid}
Y.~Shi and H.~Cheng, ``{The shadow and gamma-ray bursts of a Schwarzschild
  black hole in asymptotic safety},'' {\em Commun. Theor. Phys.}, vol.~77,
  no.~2, p.~025401, 2025.

\bibitem{neu49}
T.~Bhattacharya, S.~Habib, and E.~Mottola, ``{Gravitationally induced neutrino
  oscillation phases in static space-times},'' {\em Phys. Rev. D}, vol.~59,
  p.~067301, 1999.

\bibitem{neu53}
H.~Swami, K.~Lochan, and K.~M. Patel, ``Signature of neutrino mass hierarchy in
  gravitational lensing,'' {\em Physical Review D}, vol.~102, no.~2, p.~024043,
  2020.

\bibitem{neu47}
D.~V. Ahluwalia and C.~Burgard, ``About the interpretation of gravitationally
  induced neutrino oscillation phases,'' {\em arXiv preprint gr-qc/9606031},
  1996.

\bibitem{neu46}
D.~V. Ahluwalia and C.~Burgard, ``Gravitationally induced neutrino-oscillation
  phases,'' {\em General Relativity and Gravitation}, vol.~28, pp.~1161--1170,
  1996.

\bibitem{neu48}
Y.~Grossman and H.~J. Lipkin, ``{Flavor oscillations from a spatially localized
  source: A Simple general treatment},'' {\em Phys. Rev. D}, vol.~55,
  pp.~2760--2767, 1997.

\bibitem{neu51}
A.~Geralico and O.~Luongo, ``{Neutrino oscillations in the field of a rotating
  deformed mass},'' {\em Phys. Lett. A}, vol.~376, pp.~1239--1243, 2012.

\bibitem{neu50}
O.~Luongo and G.~V. Stagno, ``{Neutrino oscillation at the lifshitz point},''
  {\em Mod. Phys. Lett. A}, vol.~26, pp.~1257--1266, 2011.

\bibitem{neu52}
G.~Koutsoumbas and D.~Metaxas, ``{Neutrino oscillations in gravitational and
  cosmological backgrounds},'' {\em Gen. Rel. Grav.}, vol.~52, no.~10, p.~102,
  2020.

\bibitem{Shi:2023hbw}
Y.~Shi and H.~Cheng, ``{The gamma-ray burst arising from neutrino pair
  annihilation in the static and spherically symmetric black-hole-like
  wormholes},'' {\em JCAP}, vol.~10, p.~062, 2023.

\bibitem{neu54}
C.~Y. Cardall and G.~M. Fuller, ``Neutrino oscillations in curved spacetime: A
  heuristic treatment,'' {\em Physical Review D}, vol.~55, no.~12, p.~7960,
  1997.

\bibitem{neu57}
J.~Alexandre and K.~Clough, ``{Black hole interference patterns in flavor
  oscillations},'' {\em Phys. Rev. D}, vol.~98, no.~4, p.~043004, 2018.

\bibitem{neu56}
R.~M. Crocker, C.~Giunti, and D.~J. Mortlock, ``Neutrino interferometry in
  curved spacetime,'' {\em Physical Review D}, vol.~69, no.~6, p.~063008, 2004.

\bibitem{neu58}
M.~Dvornikov, ``{Spin effects in neutrino gravitational scattering},'' {\em
  Phys. Rev. D}, vol.~101, no.~5, p.~056018, 2020.

\bibitem{neu59}
J.~Zhang, M.~Liu, Z.~Liu, and S.~Yang, ``{A new touch temperature of the event
  horizon and Rindler horizon in the Kinnersley spacetime},'' {\em Eur. Phys.
  J. C}, vol.~82, no.~1, p.~1, 2022.

\bibitem{neu60}
H.~Chakrabarty, D.~Borah, A.~Abdujabbarov, D.~Malafarina, and B.~Ahmedov,
  ``Effects of gravitational lensing on neutrino oscillation in
  $\gamma$-spacetime,'' {\em The European Physical Journal C}, vol.~82, no.~1,
  p.~24, 2022.

\bibitem{Liu:2025oho}
J.-Z. Liu, S.-P. Wu, S.-W. Wei, and Y.-X. Liu, ``{Exact Black Hole Solutions in
  Bumblebee Gravity with Lightlike or Spacelike VEVS},'' 10 2025.

\bibitem{Zhu:2025fiy}
J.~Zhu and H.~Li, ``{Full Classification of Static Spherical Vacuum Solutions
  to Bumblebee Gravity with General VEVs},'' 11 2025.

\bibitem{AraujoFilho:2025zaj}
A.~A. Ara{\'u}jo~Filho, N.~Heidari, I.~P. Lobo, and V.~B. Bezerra,
  ``{Gravitational aspects of a new bumblebee black hole},'' 11 2025.

\bibitem{Salmonson:1999es}
J.~D. Salmonson and J.~R. Wilson, ``{General relativistic augmentation of
  neutrino pair annihilation energy deposition near neutron stars},'' {\em
  Astrophys. J.}, vol.~517, pp.~859--865, 1999.

\bibitem{AraujoFilho:2024mvz}
A.~A. Ara{\'u}jo~Filho, N.~Heidari, and A.~{\"O}vg{\"u}n, ``{Geodesics,
  accretion disk, gravitational lensing, time delay, and effects on neutrinos
  induced by a non-commutative black hole},'' {\em JCAP}, vol.~06, p.~062,
  2025.

\bibitem{Lambiase:2020iul}
G.~Lambiase and L.~Mastrototaro, ``{Effects of modified theories of gravity on
  neutrino pair annihilation energy deposition near neutron stars},'' {\em
  Astrophys. J.}, vol.~904, no.~1, p.~19, 2020.

\bibitem{shi2022neutrino}
Y.~Shi and H.~Cheng, ``{The neutrino pair annihilation around a massive source
  with an f(R) global monopole},'' {\em EPL}, vol.~140, no.~4, p.~49001, 2022.

\bibitem{neu18}
Y.~Nambu, S.~Noda, and Y.~Sakai, ``Wave optics in spacetimes with compact
  gravitating object,'' {\em Physical Review D}, vol.~100, no.~6, p.~064037,
  2019.

\bibitem{neu62}
Z.~Maki, M.~Nakagawa, and S.~Sakata, ``Remarks on the unified model of
  elementary particles,'' {\em Progress of Theoretical Physics}, vol.~28,
  no.~5, pp.~870--880, 1962.

\bibitem{neu63}
B.~Pontecorvo, ``Neutrino experiments and the problem of conservation of
  leptonic charge,'' {\em Sov. Phys. JETP}, vol.~26, no.~984-988, p.~165, 1968.

\bibitem{neu61}
B.~Pontecorvo, ``Inverse $ beta $ processes and nonconservation of lepton
  charge,'' {\em Zhur. Eksptl'. i Teoret. Fiz.}, vol.~34, 1958.

\bibitem{neu64}
L.~Stodolsky, ``Matter and light wave interferometry in gravitational fields,''
  {\em General Relativity and Gravitation}, vol.~11, pp.~391--405, 1979.

\bibitem{neu65}
H.~Swami, K.~Lochan, and K.~M. Patel, ``Aspects of gravitational decoherence in
  neutrino lensing,'' {\em Physical Review D}, vol.~104, no.~9, p.~095007,
  2021.

\bibitem{Liu:2024oas}
W.~Liu, D.~Wu, and J.~Wang, ``{Static neutral black holes in Kalb-Ramond
  gravity},'' {\em JCAP}, vol.~09, p.~017, 2024.

\end{thebibliography}

\end{document}